\documentclass[12pt]{article}
\usepackage{latexsym}
\usepackage{amsmath,amsbsy,amssymb}
\usepackage{delarray}

\usepackage{curves} 
\usepackage{epic}
\usepackage{eepic}
\usepackage{epsfig}

\newcommand{\dd}{\mbox{\rm d}}
\newcommand{\wg}{\wedge}
\newcommand{\gam}{\gamma}
\newcommand{\Gam}{\Gamma}
\newcommand{\nth}{\theta}
\newcommand{\bth}{{\bar \theta}}
\newcommand{\bh}{{\bar h}}

\newcommand{\dg}{\dagger}
\newcommand{\ddg}{\ddagger}
\newcommand{\tl}{\tilde}
\newcommand{\ul}{\underline}

\newcommand{\hmu}{{\hat \mu}}
\newcommand{\hnu}{{\hat \nu}}
\newcommand{\hf}{\frac{1}{2}}

\newcommand{\DD}{\mbox{\rm D}}

\newcommand{\p}{\partial}

\newcommand{\be}{\begin{equation}}
\newcommand{\bear}{\begin{eqnarray}}
\newcommand{\ear}{\end{eqnarray}}
\newcommand{\ee}{\end{equation}}

\newcommand{\lbl}{\label}
\newcommand{\bi}{\bibitem}
\newcommand{\ci}{\cite}

\newcommand{\vs}{\vspace}
\newcommand{\hs}{\hspace}

\begin{document}

\ 
\ 

\vs{0.7cm}

\begin{center}

\vs{5mm}

\baselineskip 0.6cm
{\Large \bf Quantized Fields \`a la Clifford\\ and Unification}\footnote{A chapter in the book
{\it Beyond Peaceful Coexistence; The Emergence of Space, Time and Quantum} (Edited by: Ignazio Licata,
Foreword: G. 't Hooft, World Scientific, 2016)}

\vs{4mm}

\baselineskip 0.45cm

Matej Pav\v si\v c

Jo\v zef Stefan Institute, Jamova 39, 1000 Ljubljana, Slovenia

e-mail:  matej.pavsic@ijs.si
\end{center}

\baselineskip 0.35cm
{\footnotesize It is shown that the generators of Clifford algebras
behave as creation and annihilation operators
for fermions and bosons. 
They can create extended objects, such as strings
and branes, and can induce curved metric of our spacetime.
At a fixed point, we consider the  Clifford algebra $Cl(8)$ of the 8-dimensional
phase space, and show that one quarter of the basis elements of $Cl(8)$
can represent all known particles of the first generation of the Standard model,
whereas the other three quarters are invisible to us and can thus
correspond to dark matter.}

\baselineskip 0.55cm

\section{Introduction}\label{sec1}
Quantization of a classical theory is a procedure that appears somewhat
enigmatic.  It is not a derivation in a mathematical sense.  It is a recipe of how
to replace, e.g., the classical phase space variables, satisfying the Poisson
bracket relations, with the operators satisfying the corresponding commutation
relations\,\ci{Dirac}.  What is a deeper meaning for replacement is usually
not explained, only that it works.  A quantized theory so obtained does work
and successfully describes the experimental observations of quantum phenomena.

On the other hand, there exists a very useful tool for description of geometry of
a space of arbitrary dimension and signature\,\ci{Hesteness}--\ci{Lasenby}.
This is Clifford algebra.  Its
generators are the elements that satisfy the well-known relations, namely that
the anticommutators of two generators are proportional to the components of a
symmetric metric tensor.  The space spanned by those generators is a vector
space.  It can correspond to a physical space, for instance to our usual three
dimensional space, or to the four dimensional spacetime.  The generators of a
Clifford algebra are thus basis vectors of a physical space. We will interpret
this as a space of
all possible positions that the {\it center of mass} of a physical object can
posses.  A physical object has an extension that can be described by an
effective oriented area, volume, etc.  While the center of mass position is
described by a vector, the oriented area is described by a {\it bivector}, the
oriented volume by a trivector, etc.  In general, an extended object is
described\,\ci{CastroChaos,PavsicBook,PavsicArena,PavsicMaxwellBrane,CastroPavsicRev}
 by a superposition of scalars, vectors, bivectors, trivectors, etc.,
i.e., by an element of the Clifford algebra.  The Clifford algebra associated
with an extended object is a space, called {\it Clifford space}\footnote{Here we
did not go into the mathematical subtleties that become acute when the Clifford
space is not flat but curved.  Then, strictly speaking, the Clifford space is a
{\it manifold}, such that the tangent space in any of its points is a Clifford
algebra.  If Clifford space is {\it flat}, then it is isomorphic to a Clifford
algebra.}.

Besides the Clifford algebras whose generators satisfy the anticommutation
relations, there are also the algebras whose generators satisfy commutation
relations, such that the commutators of two generators are equal to the
components of a metric, which is now {\it antisymmetric}.  The Clifford algebras
with a symmetric metric are called {\it orthogonal Clifford algebras}, whereas
the Clifford algebras with an antisymmetric metric are called {\it symplectic
Clifford algebras}\,\ci{Crumeyrole}.

We will see that symplectic basis vectors are in fact quantum mechanical
operators of bosons\,\ci{PavsicSympl,PavsicIARD12}.  The Poisson
brackets of two classical phase space
coordinates are {\it equal} to the commutators of two operators.  This is so,
because the Poisson bracket consists of the derivative and the symplectic metric
which is equal to the commutator of two symplectic basis vectors.  The
derivative acting on phase space coordinates yields the Kronecker delta and thus
eliminates them from the expression.  What remains is the commutator of the
basis vectors.

Similarly, the basis vectors of an orthogonal Clifford algebra are quantum
mechanical operators for fermions.  This becomes evident in the new basis, the
so called Witt basis.  By using the latter basis vectors and their products, one
can construct spinors.

Orthogonal and symplectic Clifford algebras can be extended to infinite
dimensional spaces\,\ci{PavsicSympl,PavsicIARD12}.  The generators
of those infinite dimensional Clifford
algebras are fermionic and bosonic field operators.  In the case of fermions, a
possible vacuum state can be the product of an infinite sequence of the
operators\,\ci{PavsicSympl,PavsicIARD12}.
  If we act on such a vacuum with an operator that does not belong to
the set of operators forming that vacuum, we obtain a ``hole" in the vacuum.
This hole behaves as a particle.  The concept of the Dirac sea, which is
nowadays considered as obsolete, is revived within the field theories based on
Clifford algebras.  But in the latter theories we do not have only one vacuum,
but many possible vacuums.  This brings new possibilities for further
development of quantum field theories and grand unification. Because the
generators of Clifford algebras are basis vectors on the one hand, and field
operators on the other hand, this opens a bridge towards quantum gravity.
Namely, the expectation values of the ``flat space" operators with respect to
suitable quantum states composed of many fermions or bosons, can give
``curved space" vectors, tangent to a manifold with non vanishing curvature.
This observation paves the road to quantum gravity.

\section{Clifford space as an extension of spacetime}\label{sec2}

Let us consider a {\it flat} space $M$ whose points are {\it possible} positions
of the center of mass $P$ of a physical object ${\cal O}$.  If the object's size
is small in comparison to the distances to surrounding objects, then we can
approximate the object ${\cal}$ with a point particle.  The squared distance
between two possible positions, with coordinates $x^\mu$ and $x^\mu + \Delta
x^\mu$, is 
\be
  \Delta s^2 = \Delta x^\mu g_{\mu \nu} \Delta x^\nu.  \lbl{2.1}
\ee
 Here index $\mu$ runs over dimensions of the space $M$, and $g_{\mu \nu}$ is
the metric tensor.  For instance, in the case in which $M$ is spacetime,
$\mu=0,1,2,3$, and $g_{\mu \nu} = \eta_{\mu \nu} = {\rm diag} (1,-1,-1,-1)$ is
the Minkowski metric.  The object ${\cal O}$ is then assumed to be extended in spacetime,
i.e., to have an extension in a 3D space and in the direction $x^0$ that
we call ``time".

There are two possible ways of taking the square root of $\Delta s^2$.

{\it Case} \ I.  
\be \Delta s = \sqrt{ \Delta x^\mu g_{\mu \nu} \Delta x^\nu}
\lbl{2.2} 
\ee

{\it Case} II.  
\be \Delta x = \Delta x^\mu \gam_\mu \lbl{2.3} 
\ee

In {\it Case} I, the square root is a {\it scalar}, i.e., the distance $\Delta
s$.

In {\it Case} II, the square root is a {\it vector} $\Delta x$, expanded in term
of the basis vectors $\gam_\mu$, satisfying the relations 
\be
  \gam_\mu \cdot \gam_\nu \equiv \frac{1}{2} (\gam_\mu \gam_\nu 
  + \gam_\nu \gam _\nu) = g_{\mu \nu} .
\lbl{2.4} 
\ee
If we write $\Delta x = \Delta x^\mu \gam_\mu = (x^\mu -
x_0^\mu) \gam_\mu$ and take $x_0^\mu = 0$, we obtain $x=x^\mu \gam_\mu$, which is
the position vector of the object's ${\cal O}$ center of mass point $P$
(Fig.\,\ref{fig1}), with $x^\mu$ being the coordinates of the point $P$.

\setlength{\unitlength}{.8mm}

\begin{figure}[h!]
\hs{3mm}
\begin{picture}(60,60)(-40,5)

\put(0,0){\includegraphics[scale=0.7]{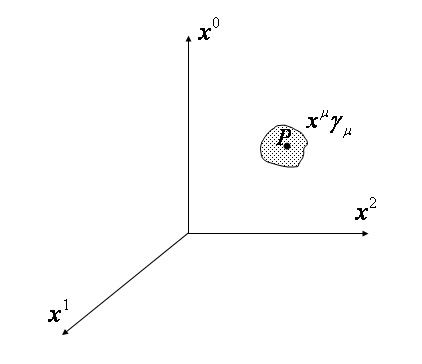}}

\end{picture}

\caption{\footnotesize The center of mass point $P$ of an extended object ${\cal O}$
is described by a vector $x^\mu \gam_\mu$.}
\lbl{fig1}
\end{figure}

In spite of being extended in spacetime and having many (practically infinitely
many) degrees of freedom, we can describe our object ${\cal O}$ by only four
coordinates $x^\mu$, the components of a vector $x=x^\mu \gam_\mu$.

The $\gam_\mu$ satisfying the anticommutation relations (\ref{2.4}) are
generators of the Clifford algebra $Cl(1,3)$.  A generic element of $Cl(1,3)$ is
a superposition 
\be
  X = \sigma {\ul 1} + x^\mu \gam_\mu + \frac{1}{2!}  x^{\mu \nu}
  \gam_\mu \wg \gam_\nu + \frac{1}{3!}  x^{\mu \nu \rho} \gam_\mu \wg \gam_\nu \wg
  \gam_\rho + \frac{1}{4!}  x^{\mu \nu \rho \sigma} \gam_\mu \wg \gam_\nu \wg
  \gam_\rho \wg \gam_\sigma,
\lbl{2.5} 
\ee
where $\gam_\mu \wg \gam_\nu$,
$\gam_\mu \wg \gam_\nu \wg \gam_\rho$ and $\gam_\mu \wg \gam_\nu \wg \gam_\rho
\wg \gam_\sigma$ are the antisymmetrized products  $\gam_\mu \gam_\nu$,
$\gam_\mu \gam_\nu \gam_\rho$, and $\gam_\mu \gam_\nu \gam_\rho \gam_\sigma$,
respectively.  They represent basis bivectors, 3-vectors and 4-vectors,
respectively.  The terms in Eq.\,(\ref{2.5}) describe a scalar, an oriented
line, area, 3-volume and 4-volume.  The antisymmetrized product of five gammas
vanishes identically in four dimensions.

A question now arises as to whether the object $X$ of Eq.\,(\ref{2.5}) can
describe an extended object in spacetime $M_4$.  We have seen that $x=x^\mu
\gam_\mu$ describes the centre of mass position.  We anticipate that
$\frac{1}{2!}  x^{\mu \nu} \gam_\mu \wg \gam_\nu$ describes an oriented area
associated with the extended object.  Suppose that our object ${\cal O}$ is a
closed string.  At first approximation its is described just by its center of
mass coordinates (Fig.\,\ref{fig2}a).  At a better approximation it is described by
the quantities $x^{\mu \nu}$, which are the projections of the oriented {\it area},
enclosed by the string, onto the coordinate planes (Fig.\,\ref{fig2}b).
If we probe the
string at a better resolution, we might find that it is not exactly a string,
but a closed membrane (Fig.\,\ref{fig3}).  The oriented volume, enclosed by this 2-dimensional
membrane is described by the quantities $X^{\mu \nu \rho}$.  At even better resolution
we could eventually see that our object ${\cal O}$ is in fact a closed
3-dimensional membrane, enclosing a 4-volume, described by $x^{\mu \nu \rho
\sigma}$.  Our object ${\cal O}$ has {\it finite extension} in the 4-dimensional
spacetime.  It is like an instanton.

\setlength{\unitlength}{.8mm}

\begin{figure}[h!]
\hs{3mm} \begin{picture}(120,40)(12,-5)

\put(47,0){\includegraphics[scale=0.6]{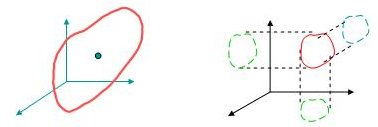}}

\put(48,20){$M_4$}
\put(74.5,19.5){$X^\mu$}
\put(73,3){$X^\mu(\xi)$}
\put(106,5){$x^1$}
\put(140,10.5){$x^2$}
\put(120,28){$x^3$}
\put(65,-6){(a)}
\put(119,-6){(b)}
\put(100,22){$X^{13}$}
\put(129,-3){$X^{12}$}
\put(146,25){$X^{23}$}

\end{picture}

\caption{\small With a closed string one can associate the center
of mass coordinates (a), and the area coordinates (b)).  }
\lbl{fig2}
\end{figure}
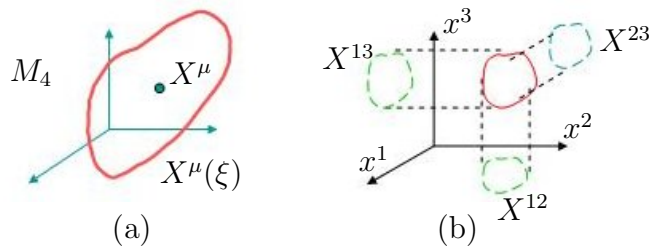 

\setlength{\unitlength}{.8mm}

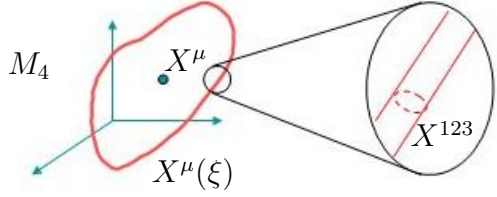
\begin{figure}[h!]
\hs{3mm} \begin{picture}(120,40)(10,-5)

\put(52,0){\includegraphics[scale=0.6]{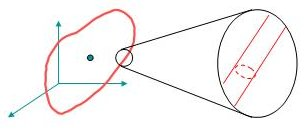}}

\put(50,20){$M_4$}
\put(76,20.5){$X^\mu$}
\put(74,2){$X^\mu(\xi)$}
\put(117,8){$X^{123}$}

\end{picture}

\caption{\small Looking with a sufficient resolution one can
detect eventual presence of volume degrees of freedom. }
\lbl{fig3}
\end{figure}

Let us now introduce a more compact notation by writing 
\be
  X= \sum_{r=0}^4
 x^{\mu_1 \mu_2 ...  \mu_r} \gam_{\mu_1 \mu_2 ...  \mu_r} \equiv x^M \gam_M,
\lbl{2.6} 
\ee
where $\gam_{\mu_1 \mu_2...\mu_r} \equiv \gam_{\mu_1} \wg \gam_{\mu_2} \wg
...\wg \gam_{\mu_r}$, and where
we now assume $\mu_1 <\mu_2 <...  <\mu_r$, so that we do not
need a factor $1/r!$.  Here $x^M$ are interpreted as quantities that describe an
extended instantonic object in $M_4$.  On the other hand, $x^M$ are coordinates
of a {\it point} in the 16-dimensional space, called {\it Clifford space} $C$.  In other
words, from the point of view of $C$, $x^M$ describe a point in $C$.

The coordinates $x^M$ of Clifford space can describe not only closed, but
also open branes. For instance, a vector $x^\mu \gam_\mu$ can denote position of
a point event with respect to the origin (Fig.\,\ref{fig1}), or it can
describe a string-like extended object (an instantonic string in spacetime).
Similarly, a bivector $x^{\mu \nu} \gam_\mu \wg \gam_\nu$ can describe
a closed string (\ref{fig2}a), or it can describe an open membrane. Whether
the coordinates $x^M \equiv x^{\mu_1 \mu_2 ... \mu_r}$ describe a closed
$r$-brane or an open $(r+1)$-brane is determined by the value of the scalar
and pseudoscalar coordinates, i.e., by $\sigma$ and ${\tl \sigma}$
(for more details see Ref.\ci{PavsicLocalTachyons}).

A continuous 1-dimensional set of points in $C$ is a curve, a {\it worldline},
described by the mapping 
\be
  x^M = X^M (\tau),
\lbl{2.7} 
\ee 
where $\tau$ is a
monotonically increasing parameter and $X^M$ embedding functions of the
worldline in $C$.  We assume that it satisfies the action principle 
\be
  I[X^M] = {\cal M} \int \dd \tau \, (G_{MN} {\dot X}^M {\dot X}^N)^{1/2},
\lbl{2.8} 
\ee 
where $G_{MN}$
is the metric in $C$, and ${\cal M}$ a constant, analogous to mass.
From the point of view of spacetime, the functions $X^M
(\tau) \equiv X^{\mu_1 \mu_2 ...  \mu_r} (\tau)$, $r=0,1,2,3,4$, describe
evolution of an extended instantonic object in spacetime.  Some examples are in
Fig.\,\ref{fig4}  (see also \ci{PavsicArena}).

\setlength{\unitlength}{.8mm}
\begin{figure}[h!]
\hs{3mm}
\begin{picture}(60,60)(-12,-4)

\put(5,-2){\includegraphics[scale=0.6]{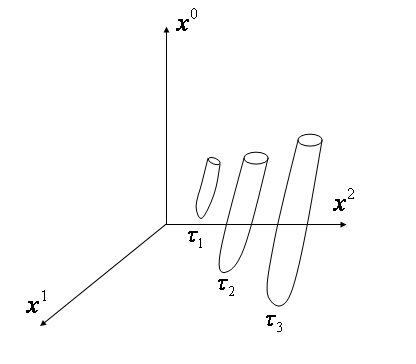}}
\put(65,0){\includegraphics[scale=0.6]{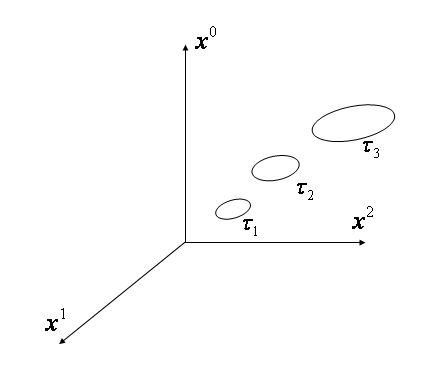}}

\put(30, -6){\rm a)}
\put(100, -6){\rm b)}

\end{picture}

\caption{\footnotesize Evolution of an instantonic cigar like (a)
and a ring like (b) extended object in spacetime.  At different
values of $\tau$, (e.g., at $\tau=\tau_1, \tau_2, \tau_3)$, we have different
extended instantonic objects that correspond to different 4D slices through
Clifford space.}
\lbl{fig4}
\end{figure} 
In this setup, there is no ``block universe" in spacetime.  There do not exist
infinitely long worldlines or worldtubes in spacetime.  Infinitely long
worldlines exist in $C$-space, and in this sense a block universe exists in
$C$-space.

The action (\ref{2.8}) is invariant under reparametrizations of $\tau$.  A
consequence is the constraint among the canonical momenta $P_M = \p L/\p {\dot
X}^M = {\cal M} {\dot X}_M/\sqrt{g_{JK}{\dot X}^J {\dot X}^K}$:  
\be
  P_M P^M - {\cal M}^2 = 0.
\lbl{2.9} 
\ee

The metric of Clifford space is given by the scalar product of two basis
vectors, 
\be
  \eta_{MN} = \gam_M^\ddg * \gam = <\gam_M^\ddg \gam_N >_0,
\lbl{2.11} 
\ee 
where ``$\ddg$'' is the operation that reverses the order of
vectors in the product $\gam_M =\gam_{\mu_1} \gam_{\mu_2}...\gam_{\mu_r}$, so
that $\gam_M^\ddg = \gam_{\mu_r} ...\gam_{\mu_2} \gam_{\mu_1}$.  The superscript
``0" denotes the scalar part of an expression.  For instance, 
\be
  <\gam_\mu \gam_\nu>_0 = \eta_{\mu \nu},
   ~~<\gam_\mu \gam_\nu \gam_\alpha>_0 = 0,~~
<\gam_\mu \gam_\nu \gam_\alpha \gam_\beta>_0 = \eta_{\mu \beta} \eta_{\nu
\alpha} - \eta_{\mu \alpha} \eta_{\nu \beta}.
\lbl{2.12} 
\ee
 So we obtain 
\be
  \eta_{MN} = {\rm diag} (1,1,1,1,1,1,1,1,-1,-1,-1,-1,-1,-1,-1,-1),
\lbl{2.12a}
\ee
which means that the signature of $C$-sapce is $(++++++++--------)$, or
shortly, $(8,8)$.

The quadratic form reads 
\bear
  X^\ddg * X &=& \eta_{MN} x^M x^N \nonumber\\
  &=& \sigma^2+\eta_{\mu \nu} x^\mu x^\nu + (\eta_{\mu \beta} \eta_{\nu \alpha}
   - \eta_{\mu \alpha} \eta_{\nu \beta}) x^{\mu \alpha} x^{\nu \beta}
    + \eta_{\mu \nu} {\tl x}^\mu {\tl x}^\nu - {\tl \sigma}^2 \nonumber\\
 &=& \eta_{{\hat \mu}{\hat \nu}} x^\hmu x^\hnu
+ \sigma^2 - {\tl \sigma}^2, 
\lbl{2.13} 
\ear
where $x^\hmu = (x^\mu,x^{\mu \nu},{\tl x}^\mu)$, with
${\tl x}^\mu \equiv \frac{1}{3!} {\epsilon^\mu}_{\nu \rho \sigma}
x^{\nu \rho \sigma}$ being the pseudoscalar coordinates, whereas $\sigma$
is the scalar and ${\tl \sigma}
\equiv \frac{1}{4!} \epsilon_{\mu \nu \rho \sigma} x^{\mu \nu \rho \sigma}$
the pseudoscalar coordinate in $C$-space.

Upon quantization, $P_M$ become operators $P_M = - i \p/\p x^M$, and the
constraint (\ref{2.9}) becomes the Klein-Gordon equation in $C$-space:  
\be
  (\p_M \p^M + {\cal M}^2) \Psi (x^M) = 0.  
\lbl{2.10} 
\ee 
In the new coordinates,
\be
  s = \frac{1}{2}(\sigma+{\tl \sigma})\;
  ~~~~~ \lambda =\frac{1}{2}(\sigma-{\tl \sigma}), 
\lbl{2.14} 
\ee 
in which the quadratic form is 
\be
  X^\ddg * X= \eta_{\hmu \hnu} x^\hmu x^\hnu - 2 s \lambda, 
\lbl{2.15} 
\ee 
the Klein-Gordon
equation reads 
\be
  \eta^{\hmu \hmu} \p_\hmu \p_\hnu \phi - 2 \p_s \p_\lambda \phi = 0.  
\lbl{2.16} 
\ee
 If we take the ansatz 
 \be
  \phi(x^\hmu,s,\lambda) = {\rm e}^{i \Lambda \lambda} \psi(s,x^\hmu), 
\lbl{2.17} 
\ee 
then Eq.\,(\ref{2.16}) becomes\,\ci{PavsicIARD10} 
\be
  \eta^{\hmu \hmu} \p_\hmu \p_\hnu \phi - 2 i \Lambda \p_s \phi = 0, 
\lbl{2.18} 
\ee 
i.e., 
\be
  i \frac{\p \psi}{\p s} =
  \frac{1}{2 \Lambda} \eta^{\hmu \hmu} \p_\hmu \p_\hnu \psi .  
\lbl{2.19} 
\ee 
This
is the generalized {\it Stueckelberg equation}.  It is like the Schr\"odinger
equation, but it describes the evolution of the wave function $\psi (s,x^\hmu)$
in the 14-dimensional space whose points are described by coordinates $x^\hmu$.
The evolution parameter is $s$.

A remarkable feature of this setup is that the evolution parameter has a clear
physical meaning:  it is given in terms of the scalar, $\sigma$, and the
pseudoscalar, ${\tl \sigma}$, coordinate according to Eq.\,(\ref{2.14}).  The
latter quantities, as shown before, are given by a configuration of the object,
sampled in terms of the coordinates $X^M$ of the Clifford space $C$.

The wave function $\psi(s,x^\hmu)$ is the probability amplitude that at a given
value of the evolution parameter $s$ we will find an instantonic extended object
with coordinates $x^\hmu$.

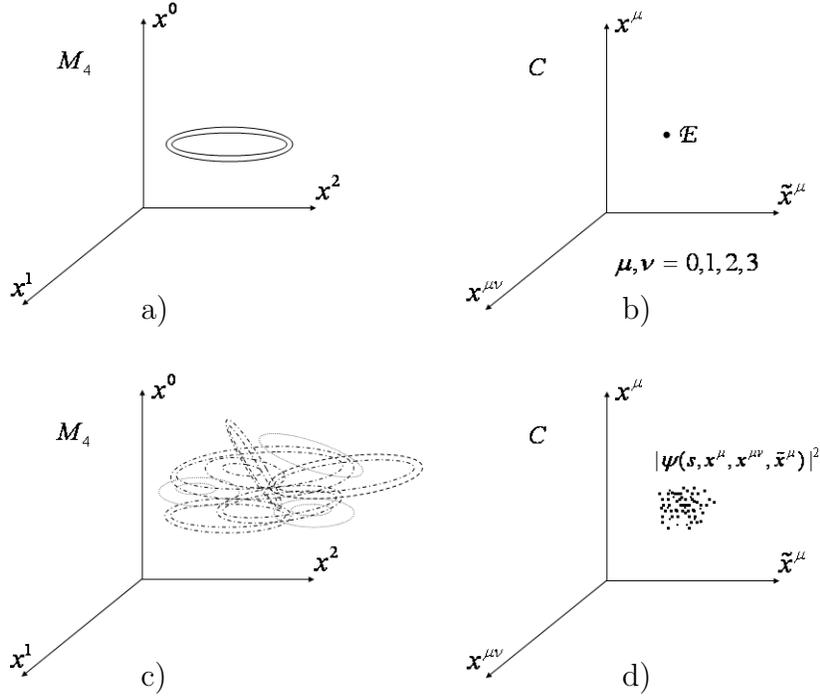
\begin{figure}[h!]
\hs{3mm}
\begin{picture}(120,120)(-18,0)

\put(-2,60){\includegraphics[scale=0.6]{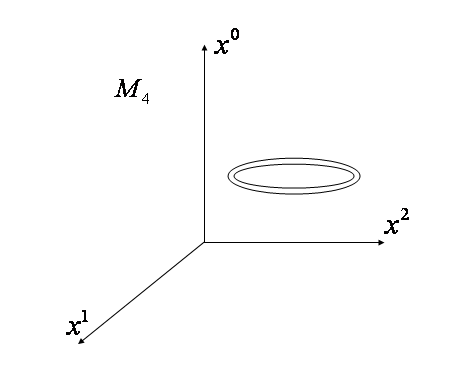}}
\put(77,60){\includegraphics[scale=0.6]{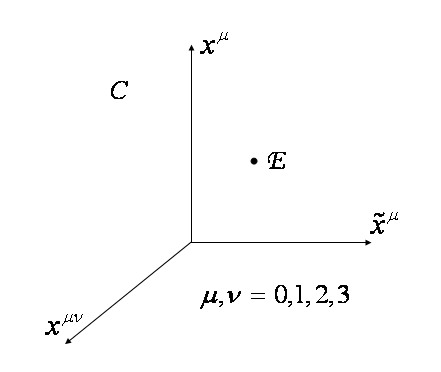}}
\put(0,0){\includegraphics[scale=0.6]{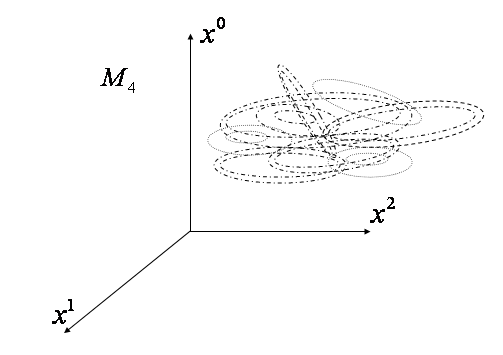}}
\put(79,-2){\includegraphics[scale=0.6]{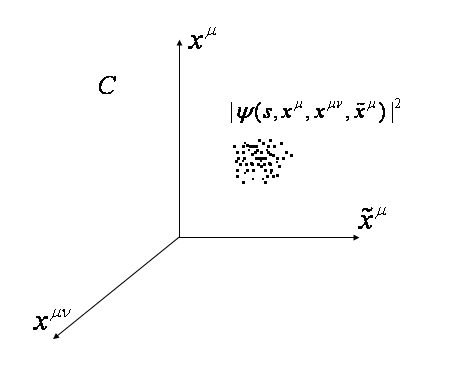}}

\put(30, 63){\rm a)}
\put(110, 63){\rm b)}
\put(30, 2){\rm c)}
\put(110, 2){\rm d)}

\end{picture}

\caption{\footnotesize Extended instantonic object in spacetime (a) is represented by a point in
$C$-space (b).  Quantum mechanically, the extended object is blurred (c).  In
$C$-space, we have a blurred point, i.e., a ``cloud" of points occurring with
probability density $|\psi (s,x^\hmu)|^2$.
}
\lbl{fig5}
\end{figure}

This is illustrated in Fig.\,\ref{fig5}.  In principle all points of $C$-space are
possible in the sense that we can find there an instantonic extended object.
A wave packet determines a subset of point of $C$ that are more probable to
``host" the occurrence of an instantonic object (an event in $C$).
The wave function determines
the probability amplitude over the points of $C$.  Its square determines the
probability density.  From the point of view of spacetime, wave function
determines which instantonic extended objects are more likely to occur.  It
determines the probability amplitude, and its square the probability density of
occurrence of a given instantonic extended object.  The probability amplitude
$\psi$ is different at different values of the evolution parameter $s$.  In
other words, $\psi$ changes (evolves) with $s$.

Instead of one extended object, described by $x^M$, we can consider several or
many extended objects, described by $x^{iM}$, $i=1,2,...,n$.  They form an
instantonic configuration $\{{\cal O}^i\} =\{{\cal O}^i\}, i = 1,2,...,n$.
The space of
all possible instantonic configurations will be called configuration space
${\cal C}$.  The infinitesimal distance between two configurations, i.e.,
between two points in ${\cal C}$, is 
\be
  \dd S^2 = \eta_{(iM)(jN)} \dd x^{iM} \dd x^{jN} , 
\lbl{2.20} 
\ee 
where $\eta_{(iM)(jN)} = \delta_{ij} \eta_{MN}$ is
the metric of a {\it flat} configuration space.

We will assume that the Klein-Gordon equation (\ref{2.10}) can be generalized so
to hold for the wave function $\psi(x^{iM})$ in the space of instantonic
configurations $\{{\cal O}^i\}$:  
\be
  \left (\eta^{(iM)(jN)} \p_{iM} \p_{jN} + {\cal K}^2 \right ) \phi(x^{iM}) 
  = 0~, ~~~~~\p_{iM} \equiv \frac{\p}{\p X^{iM}}  
\lbl{2.21} 
\ee

Let us choose a particular extended object, ${\cal O}^1$, with coordinates
$x^{1M} \equiv x^M = (\sigma,x^\mu,x^{\mu \nu}, {\tl x}^\mu, {\tl \sigma})$. The
coordinates of the remaining extended objects within the configuration are
$x^{2M}$, $x^{3M}$,\,...\,.  Let us denote them $x^{{\bar i}M}$, ${\bar
i}=2,3,...,N$.  Following the same procedure as in
Eqs.\,(\ref{2.14})--(\ref{2.19}), we define $s$ and $\lambda$ according to
(\ref{2.14}) to the first object.  We have thus split the coordinates $x^{iM}$
of the configuration according to 
\be
  x^{iM} = (s,\lambda, x^\hmu,x^{{\bar i} M}) = (s,\lambda,x^{\bar M}), 
\lbl{2.22} 
\ee 
where $x^{\bar M}= (x^\hmu,x^{{\bar
i}M})$.  By taking the ansatz 
\be
  \phi (s,\lambda, x^{\bar M}) = {\rm e}^{i \Lambda \lambda} \psi(s,x^{\bar M}), 
\lbl{2.23} 
\ee 
Eq.\,(\ref{2.21}) becomes
\be
  \eta^{{\bar M}{\bar N}} \p_{\bar M} \p_{\bar N} \psi 
  - 2 i \Lambda \p_s \psi = 0, 
\lbl{2.24} 
\ee 
i.e., 
\be
  i \frac{\p}{\p s} \psi = \frac{1}{2 \Lambda} \eta^{{\bar M}{\bar N}} \p_{\bar M} 
  \p_{\bar N} \psi .  
\lbl{2.25} 
\ee 
Eq.\,(\ref{2.25})
describes evolution of a configuration composed of a system of instantonic
extended objects.

The evolution parameter $s$ is given by the configuration itself (in the above
example by one of its parts), and it distinguishes one instantonic configuration
from another instantonic configuration.  So we have a continuous family of
instantonic configurations, evolving with $s$.  Here, ``instantonic
configuration" or ``instantonic extended object'' is a generalization of the
concept of ``event", associated with a point in spacetime.  An event, by
definition is ``instantonic" as well, because it occurs at one particular point
in spacetime.

A configuration can be very complicated and self-referential, and thus being a
record of the configurations at earlier values of $s$.  In this respect this
approach resembles that by Barbour\ci{BarbourEnd}, who considered ``time
capsules" with memory of the past.  As a model, he considered a triangleland,
whose configurations are triangles.  Instead of triangleland, we consider here
the Clifford space, in which configurations are modeled by oriented $r$-volumes
($r=0,1,2,3,4$) in spacetime.  In this respect our model differs from Barbour's
model, in which the triangles are in 3-dimensional space.  Instead of
3-dimensional space, I consider a 4-dimensional space with signature $(+---)$.
During the development of physics it was recognized that a 3-dimensional space
is not suitable for formulation of the theory describing the physical phenomena,
such as electromagnetism and moving objects.  In other words, the theory of
relativity requires 4-dimensional space, with an extra dimension $x^0$, whose
signature is opposite to the signature of three spatial dimensions.  The fourth
dimension was identified with time, $x^0 \equiv t$.  Such identification, though
historically very useful, has turned out to be
misleading \,\ci{Horwitz,PavsicEvolTime,FanchiBook},\ci{PavsicBook}.
In fact, $x^0$ is not the true time, it is just a
coordinate of the fourth dimension.  The evolution time is something else.  In
the Stueckelberg theory\,\ci{Horwitz,PavsicEvolTime,FanchiBook},\ci{PavsicBook}
its origin remains unexplained.  In
the approach with Clifford space, the evolution time (evolution parameter) is
$s=(\sigma + {\tl \sigma})/2$, i.e., a superposition of the {\it scalar}
coordinate, $\sigma$, and the {\it pseudoscalar} coordinate, ${\tl \sigma}$.
This is the parameter
that distinguishes configurations within a 1-dimensional family.  In principle,
the configurations can be very complicated and self-referential, including
conscious experiences of an observer.  Thus $s$ distinguishes different
conscious experiences of an observer\,\ci{PavsicEvolTime,PavsicBook};
it is the time experienced by a conscious
observer.  A wave function $\phi(x^{iM}) = {\rm e}^{i \Lambda} \psi(s,x^{\bar
M})$ ``selects" in the vast space ${\cal C}$ of all possible configurations
$x^{iM}$ a subspace ${\cal S} \in {\cal C}$ of configurations.  More precisely,
$\phi$ assigns a probability density over the points of ${\cal C}$, so that some
points are more likely to be experienced by an observer than the other points.
In particular, $\phi(x^{iM})$ can be a  localized wave packet
evolving along $s$. For instance, such a wave packet
can be localized around a worldline $x^{iM} = X_0^{i{M}} (s)$ in ${\cal C}$,
which from the point of view of $M_4$, is a succession (evolution) of
configurations $X_0^{{\bar M}}$ at different values of the parameter $s$.
If configurations are complicated and include the external world and an
observer's brain, such wave packet $\psi(s,x^{\bar M})$ determines the
evolution of conscious experiences of an observer coupled by his sense
organs to the external world.

The distinction between the evolution time $\tau$ and
the coordinates $x^0$ in the wave function $\psi(s,x^{\bar M})$, can help
in clarifying the well known Libet experiment\,\ci{Libet}. The latter
experiment seemingly
demonstrates that we have no ``free will", because shortly before we are
conscious of a decision, our brain already made the decision. This experiment
is not in conflict with free will, if besides the theory described above, we
as well invoke the Everett many worlds interpretation of quantum
mechanics\,\ci{Everett}, and the considerations exposed in Ref.\,\ci{PavsicBook}.
Further elaboration of this important implication of the Stueckelberg and
Everett theory is beyond the scope of this chapter. But anyone with a
background in those theories can do it after some thinking. An interested reader
can do it as an exercise.

\section{Generators of Clifford algebras as quantum mechanical
operators}\lbl{sec3}

\subsection{Orthogonal and symplectic Clifford algebras}

After having exposed a broader context of the role of Clifford algebras in
physics, let me now turn to a specific case and consider the role of Clifford
algebras in quantization.  The inner product of generators of Clifford algebra
gives the metric.  We distinguish two cases:

\ (i) If metric is {\it symmetric}, then the inner product is given by the {\it
anticommutator} of generators; this is the case of an {\it orthogonal Clifford
algebra}:  
\be
  \frac{1}{2} \lbrace \gam_a,\gam_b \} \equiv \gam_a \cdot \gam_b =g_{ab}.  
\lbl{3.1} 
\ee

(ii) If metric is {\it antisymmetric}, then the inner product is given by the
commutator of generators; this is the case of a {\it symplectic Clifford
algebra}:  
\be
  \frac{1}{2} [q_a,q_b] \equiv q_a \wg q_b = J_{ab} .  
\lbl{3.2}
\ee

Here $q_a$ are the symplectic basis vectors that span a {\it symplectic space},
whose points are associated with symplectic vectors\,\ci{PavsicSympl} 
\be
  z = z^a q_a .  
\lbl{3.3}
\ee
 Here $z^a$ are commuting phase space coordinates, 
\be
  z^a z^b - z^b z^a = 0.
\lbl{3.3a} 
\ee
 An example of symplectic space in physics is {\it phase space},
whose points are coordinates and momenta of a particle:  
\be
  z^a = (x^\mu,p^\mu) \equiv (x^\mu,{\bar x}^\mu) \equiv (x^\mu,x^{\bar \mu}).  
\lbl{3.4} 
\ee
The corresponding basis vectors then split according to 
\be
  q_a =
(q_\mu^{(x)},q_\mu^{(p)}) \equiv (q_\mu,{\bar q}_\mu) \equiv (q_\mu,q_{\bar
\mu}),~~~~~~~ \mu = 1,2,...,n, 
\lbl{3.5} 
\ee 
and the relation (\ref{3.2}) becomes 
\bear
&&\mbox{$\frac{1}{2}$} [q_\mu^{(x)},q_\nu^{(p)}] \equiv \frac{1}{2} [q_\mu,q_{\bar \nu}] =
J_{\mu {\bar \nu}} = g_{\mu \nu},\nonumber\\
&& \hs{0.5cm} [q_\mu^{(x)},q_\nu^{(x)}] =
0,~~~~~  [q_\mu^{(p)},q_\nu^{(p)}] = 0, 
\lbl{3.6} 
\ear
where we have set 
\be
 J_{ab} = \begin{pmatrix} 0 & g_{\mu \nu}\\ -g_{\mu \nu} & 0
\end{pmatrix} .  
\lbl{3.7} 
\ee
 Here, depending on the case considered, $g_{\mu
\nu}$ is the euclidean, $g_{\mu \nu}= \delta_{\mu \nu}$, $\mu ,\nu = 1,2,...,n$,
or the Minkowski metric, $g_{\mu \nu} = \eta_{\mu \nu}$.  In the latter case we
have $\mu,\nu=0,1,2,...,n-1$.

We see that (\ref{3.6}) are just the Heisenberg commutation relations for
coordinate and momentum operators, identified as\footnote{ We insert factor $i$
in order to make the operator ${\hat p}_\mu$ hermitian.}
\be
  {\hat x}_\mu =
  \frac{1}{\sqrt{2}} q_\mu^{(x)}~;~~~~~~~~ {\hat p}_\mu = \frac{i}{\sqrt{2}}
  q_\mu^{(p)} .  
\lbl{3.7a} 
\ee 
Then we have 
\be
  [{\hat x}_\mu,{\hat p}_\nu] = i g_{\mu \nu} ~,~~~~~ 
  [{\hat x}_\mu,{\hat x}_\nu] = 0 ~,~~~~~ [{\hat p}_\mu,{\hat p}_\nu] = 0 .  
\lbl{3.8} 
\ee

Instead of a symplectic vector $z=z^a q_a$, let us now consider another
symplectic vector, namely 
\be
  F= \frac{\p f}{\p z^a} q^a , 
\lbl{3.9} 
\ee 
where $f = f(z)$ is a function of position in phase space. The wedge product
of two such vectors is 
\be
  F \wg G = \frac{\p f}{\p z^a} q^a \wg  q^b \frac{\p g}{\p z^b} =
  \frac{\p f}{\p z^a} J^{ab} \frac{\p g}{\p z^b} , 
\lbl{3.10} 
\ee 
where in the
last step we used the analog of Eq.\,(\ref{3.2}) for the reciprocal quantities
$q^a = J^{ab} q_b$, where $J^{ab}$ is the inverse of $J_{ab}$.

Eq.\,(\ref{3.10}) is equal to the Poisson bracket of two phase space functions.
namely, using (\ref{3.4}) and (\ref{3.7}), we have 
\be
  \frac{\p f}{\p z^a} J^{ab} \frac{\p g}{\p z^b} 
  = \frac{\p f}{\p x^\mu} \eta^{\mu \nu} \frac{\p g}{\p p^\nu}
  - \frac{\p f}{\p p^\mu} \eta^{\mu \nu} \frac{\p g}{\p x^\nu} \equiv
  \lbrace f,g \rbrace_{PB} .  
\lbl{3.11} 
\ee 
In particular, if
 \be
 f = z^c~,~~~~~g = z^d, 
\lbl{3.12} 
\ee 
Eqs.\,(\ref{3.10}),(\ref{3.11}) give 
\be
  q^a \wg q^b = J^{ab} = \lbrace z^a,z^b \rbrace_{PB} .  
\lbl{3.13} 
\ee

We see that the Heisenberg commutation relations for operators ${\hat x}^\mu$,
${\hat p}^\mu$ are obtained automatically, if we express the Poisson bracket
relations in terms of the wedge product of the symplectic vectors 
\be
  F= \frac{\p f}{\p z^a} q^a = \frac{\p f}{\p x^\mu} q^\mu_{(x)} 
  +\frac{\p f}{\p p^\mu} q^\mu_{(p)} ~~~{\rm and}~~~ 
  G= \frac{\p g}{\p z^a} = \frac{\p g}{\p x^\mu}
 q^\mu_{(x)} +\frac{\p f}{\p p^\mu} q^\mu_{(p)} 
\lbl{3.13a} 
\ee 
By having taken
into account not only the coordinates and functions in a symplectic space, but
also corresponding basis vectors, we have found that basis vectors are in fact
quantum mechanical operators\,\ci{PavsicSympl}.
  Moreover, the Poisson bracket between classical
phase space variable, $\{z^a,z^b\}_{PB}$, is {\it equal} to the commutator,
$\frac{1}{2} [q^a,q^b] =q^a \wg q^b$, of vectors (i.e., of operators) $q^a$ and
$q^b$\,\ci{PavsicSympl}.  According to this picture, quantum operators
are already present in the
classical symplectic form, if we write the symplectic metric as the inner
product of symplectic basis vectors.  The latter vectors are just the quantum
mechanical operators.

Analogous procedure can be performed with orthogonal Clifford algebras.  Then a
point in phase space can be described as a vector 
\be
  \lambda = \lambda^a \gam_a, 
\lbl{3.14} 
\ee 
where $\lambda^a$ are anticommuting phase space coordinates,
\be
  \lambda^a \lambda^b + \lambda^b \lambda^a = 0,
\lbl{3.14a}
\ee
and $\gam_a$ basis vectors, satisfying Eq.\,(\ref{3.1}).  If we split the
vectors $\gam_a$ and the metric $\gam_{ab}$ according to 
\be
  \gam_a = (\gam_\mu, {\bar \gam}_\mu) , ~~~~\mu=0,1,2,...,n-1, 
\lbl{3.15} 
\ee
\be
  g_{ab} = \begin{pmatrix}
                g_{\mu \nu} & 0 \\
                  0 & g_{\mu \nu}
            \end{pmatrix}
\lbl{3.16a}
\ee             
and introduce a new basis, the so
called {\it Witt basis}, 
\bear
  &&\theta_\mu = \frac{1}{\sqrt{2}} (\gam^\mu + i {\bar \gam}_\mu ),\nonumber\\ 
  &&{\bar \theta}_\mu = \frac{1}{\sqrt{2}}
  (\gam^\mu - i {\bar \gam}_\mu ), \lbl{3.15a}
\ear
then the Clifford algebra relations (\ref{3.1}) become 
\bear
  \theta_\mu \cdot {\bar \theta}_\nu \equiv 
  \frac{1}{2} \,(\theta _\mu {\bar \theta}_\nu 
  + {\bar \theta}_\nu \theta_\mu) = \eta_{\mu \nu}, \nonumber \\
 ~~~~~ \theta_\mu \cdot \theta_\nu = 0,~~~~~
 {\bar \theta}_\mu \cdot {\bar \theta}_\nu = 0.
\lbl{3.16}
\ear
These are the anticommutation relations for fermionic creation and annihilation
operators.

Let us now introduce functions ${\tl f} (\lambda)$ and ${\tl g}(\lambda)$, and
consider the vectors 
\be
  {\tl F} = \frac{\p {\tl f}}{\p \lambda^a} \gam^a ,
  ~~~~~~ {\tl g} = \frac{\p {\tl g}}{\p \lambda^a} \gam^a 
\lbl{3.17} 
\ee 
The dot product of those vectors is 
\be
  {\tl F} \cdot {\tl G} = \frac{\p {\tl f}}{\p \lambda^a} \gam^a \cdot 
  \frac{\p {\tl g}}{\p \lambda^b} \gam^b
   = \frac{\p {\tl f}}{\p \lambda^a} g^{ab} \frac{\p {\tl g}}{\p \lambda^b} 
   =\lbrace {\tl f},{\tl g} \rbrace_{PB} ,
\lbl{3.18} 
\ee
where $g^{ab} = \gam^a \cdot \gam^b$ is the inverse of $g_{ab}$.

Eq.\,(\ref{3.18}) shows that the dot product, which in the orthogonal case
corresponds to the inner product, is equal to the Poisson bracket of two phase
space functions, now composed with the symmetric metric $g^{ab}$.

If 
\be
  {\tl f} = \lambda^c~,~~~~~~~{\tl g} = \lambda^d, 
\lbl{3.19} 
\ee
Eq.\,(\ref{3.18}) gives
 \be
  {\tl F} \cdot {\tl G} = \gam^c \cdot \gam^d = g^{cd},
\lbl{3.20}
\ee
which in the Witt basis read as the fermionic anticommutation relations
(\ref{3.16}). This means that the
Poisson bracket between the (classical) phase space variables $\lambda^a$,
$\lambda^b$ is equal to the anticommutator of the ``operators'' $\gam^a$ and
$\gam^b$:  
\be
  \lbrace \lambda^a,\lambda^b \rbrace_{PB} = \frac{1}{2} \{\gam^a,\gam^b \} 
  =g^{ab} .  
\lbl{3.21} 
\ee 
Again we have that the basis
vectors behave as quantum mechanical operators.

\subsection{Equations of motion for a particle's coordinates and the
corresponding basis vectors}

We will now consider\,\ci{PavsicSympl} a point particle, described by the phase space action 
\be I
= \frac{1}{2} \int \dd \tau \, ( {\dot x}^a J_{ab} z^b + z^a K_{ab} z^b ),
\lbl{3.22} 
\ee
 where 
\be
  \frac{1}{2}z^a K_{ab} z^b = H 
\lbl{3.23} 
\ee is the Hamiltonian, the quantity $K_{ab}$ being a symmetric
$2n \times 2n$ matrix.

Variation of the action (\ref{3.22}) with respect to $z^a$ gives 
\be
  {\dot z}^a = J^{ab}\, \frac{\p H}{{\p z}^b}, 
\lbl{3.24} 
\ee 
which are the Hamilton equations of motion.

A solution of equation (\ref{3.24}) is a trajectory $z$ in phase space.  We can
consider a trajectory as an infinite dimensional vector with components $z^a
(\tau) \equiv z^{a(\tau)}$.  Here $a(\tau)$ is the index that denotes
components; it is a double index, with $a$ being a discrete index, and $(\tau)$
a continuous one.  Corresponding basis vectors are $q_a (\tau) \equiv
q_{a(\tau)}$, and they satisfy the relations 
\be
  q_{a(\tau)} \wg q_{b(\tau)} = J_{a(\tau) b(\tau)} 
  = J_{ab} \delta (\tau-\tau'), 
\lbl{3.25}
\ee 
which are an extension of the relations (\ref{3.2}) to our infinite dimensional
case.

A trajectory is thus described by the vector 
\be
  z= z^{a(\tau)} q_{a(\tau)} \equiv \int \dd \tau z^a (\tau) q_a (\tau).  
\lbl{3.26} 
\ee
 The phase space velocity vector is 
\be
  v = {\dot z}^{a(\tau)} q_{a(\tau)} = - z^{a(\tau)} {\dot q}_{a(\tau)},
\lbl{3.27} 
\ee 
where we have assumed that the ``surface" term vanishes:  
\be
  v= \int \dd \tau {\dot z}^a (\tau) q_a (\tau) = - \int \dd \tau
  z^a (\tau) {\dot q}_a (\tau) + z^a (\tau) q_a (\tau)\Bigl\vert_{\tau_1}^{\tau_2}.
\lbl{3.28} 
\ee 
The last term vanishes if $z^a (\tau_2) q_a(\tau_2)=z^a (\tau_1) q_a (\tau_1)$.

The action (\ref{3.22}) can be written as 
\be
  I= \frac{1}{2} \left ( {\dot z}^{a(\tau)} J_{a(\tau) b(\tau)} z^{b(\tau')}
   +z^{a(\tau)} K_{a(\tau) b(\tau')} z^{b (\tau')} \right ), 
\lbl{3.29} 
\ee
 where $ J_{a(\tau) b(\tau)}$ is given in
Eq.\,(\ref{3.25}), and
\be
  K_{a(\tau) b(\tau')}= K_{ab} \delta(\tau-\tau').
\lbl{3.30} 
\ee
 The corresponding equations of motion are 
\be
  {\dot z}^{a(\tau)}
  = J^{a(\tau) c(\tau'')} K_{c(\tau'')b(\tau')} z^{b (\tau')}.  
\lbl{3.31} 
\ee
Multiplying both sides of the latter equation by $q_{a(\tau)}$, we obtain 
\be
 {\dot z}^{a(\tau)} q_{a(\tau)} =-q^{a(\tau)} K_{a(\tau) b(\tau')} z^{b(\tau')}.
\lbl{3.32} 
\ee
 We have raised the index by $J^{a(\tau) c(\tau'')}$ and taken
into account that $J^{a(\tau) c(\tau'')}= - J^{c(\tau'') a(\tau)}$.
Eq.\,(\ref{3.32}) is just Eq.\,(\ref{3.31}), expressed in terms of the basis
vectors.  Both equations are equivalent.

Using the relation (\ref{3.27}) in Eq.\,(\ref{3.32}), we obtain 
\be
  z^{b(\tau')} {\dot q}_{b(\tau')} 
  = q^{a(\tau)} K_{a(\tau) b(\tau')} z^{b(\tau')} 
\lbl{3.33}
\ee
Apart from the surface term that we have neglected in Eq.\,(\ref{3.28}), 
the last equation, (\ref{3.33}), is equivalent to the classical equation of motion
(\ref{3.24}), only the $\tau$-dependence has been switched from the components
to the basis vectors. 

A curious thing happens if we assume that Eq.\,(\ref{3.33}) holds for an
arbitrary trajectory (Fig.\,\ref{fig6}). 

\begin{figure}[h!]
\hs{3mm}
\begin{picture}(60,60)(-50,5)

\put(-10,0){\includegraphics[scale=0.65]{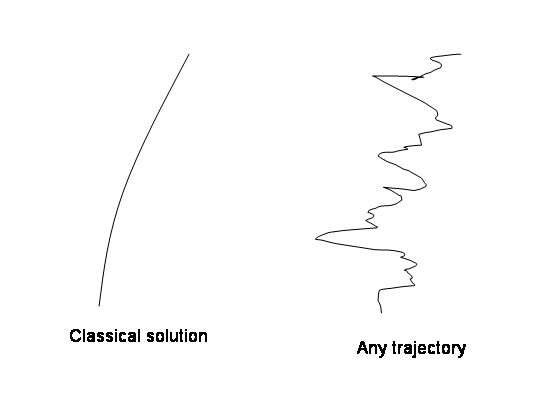}}

\end{picture}

\caption{\footnotesize If the operator equations of motion (\ref{3.33})
hold for any path $z^a(\tau)$ this means that coordinates and momenta
are undetermined.}

\lbl{fig6}
\end{figure}

Then, instead of (\ref{3.33}), we can write

\be
  {\dot q}_{b(\tau')} = q^{a(\tau)} K_{a(\tau) b(\tau')} 
\lbl{3.34} 
\ee
Inserting into the latter equation the explicit expression (\ref{3.30}) for
$K_{a(\tau) b(\tau')}$ and writing ${\dot q}_{b(\tau)}={\dot q}_b (\tau)$,
$q^{a(\tau)}= q^a (\tau)$, we obtain 
\be {\dot q}_a (\tau) = K_{ab} q^b (\tau).  
\lbl{3.35} 
\ee
This can be written as 
\be {\dot q}_a = [q_a,{\hat H}],
\lbl{3.36} 
\ee
where 
\be {\hat H} = \frac{1}{2} q^a K_{ab} q^b ,
\lbl{3.37} 
\ee 
is the Hamilton operator, satisfying 
\be [q_a,{\hat H}] = K_{ab} q^b .  
\lbl{3.38} 
\ee

Starting from the classical action (\ref{3.22}), we have arrived at the
Heisenberg equations of motion (\ref{3.36}) for the basis vectors $q_a$.  On the
way we have made a crucial assumption that the particle does not follow a
trajectory $z^a(\tau)$ determined by the classical equations of motion, but that
it can follow any trajectory.  By the latter assumption we have passed from the
classical to the quantized theory.  We have thus found yet another way of
performing quantization of a classical theory, Our assumption that a trajectory
(a path) can be arbitrary, corresponds to that by Feynman path integrals.  In
our procedure we have shown how such an assumption of arbitrary path leads to
the Heisenberg equations of motion for operators.

\subsection{Supersymmetrization of the action}

The action (\ref{3.22}) can be generalized\,\ci{PavsicSympl}
so to contain not only a symplectic,
but also an orthogonal part.  For this purpose, we introduce the generalized
vector space whose elements are 
\be z= z^A q_A, 
\lbl{3.39} 
\ee
 where 
\bear
 &&z^A =
(z^a,\lambda^a)~,~~~~~~~z^a = (x^\mu,{\bar x}^\mu)~,~~~~~~~~~\lambda^a = (\lambda ^\mu,
{\bar \lambda}^\mu) \lbl{3.40}\\ 
   &&\hs{3cm} {symplectic~part} \hs {1cm}
{orthogonal~ part} \nonumber 
\ear 
are coordinates, and 
\bear
   &&q_A =
(q_a,\gam_a)~,~~~~~~~~~q_a=(q_\mu,{\bar q}_\mu)~, ~~~~~~~~~\gam_a=(\gam_\mu,{\bar
\gam}_\mu) \lbl{3.41}\\ 
 &&\hs{3cm} {symplectic~part} \hs {1cm} {orthogonal~ part} \nonumber 
\ear 
are basis vectors.  The metric is 
\be \langle
q_A q_B \rangle_0 =G_{AB} =
 \begin{pmatrix} J_{ab} & 0 \\ 0 & g_{ab}
\end{pmatrix}, 
\lbl{3.42} 
\ee
 where $J_{ab} =- J_{ba}$ and $g_{ab} = g_{ba}$.

Let us consider a particle moving in such space.  Its worldline is 
\be z^A =
Z^A(\tau).  
\lbl{3.43} 
\ee 
An example of a possible action is 
\be I= \frac{1}{2}
\int \dd \tau \, {\dot Z}^A G_{AB} Z^B + {\rm interaction ~ terms}.  
\lbl{3.45}
\ee 
Using (\ref{3.40})--(\ref{3.42}), the latter action can be split as
 \bear
 I &=& \frac{1}{2} \int \dd \tau \, ({\dot z}^a J_{ab} z^b + {\dot \lambda}^a
  g_{ab} \lambda ) + {\rm interaction~terms}\nonumber \\
  &=&\frac{1}{2} \int \dd
  \tau \, ({\dot x}^\mu \eta_{\mu \nu} {\bar x}^\nu - {\dot {\bar x}}^\mu
 \eta_{\mu \nu} x^\nu + {\dot {\lambda}}^\mu \eta_{\mu \nu} \lambda^\nu + {\dot
  {\bar \lambda}} \eta_{\mu \nu} {\bar \lambda}^\nu )\nonumber\\
  &&\hs{1cm} + \,{\rm interaction~terms}
\lbl{3.46}
\ear
Here $z^a$ are commuting, and $\lambda^a$ anticommuting
(Grassmann) coordinates.  The canonical momenta are 
\bear
&&p_\mu^{(x)} = \frac{\p L}{\p {\dot x}^\mu}=\hf \eta_{\mu \nu} {\dot x}^\nu~,
 ~~~~~p_\mu^{({\bar x})} = \frac{\p L}{\p {\dot {\bar x}}^\mu} = -\hf \eta_{\mu
 \nu} x^\nu \nonumber\\
 &&p_\mu^{(\lambda)} = \frac{\p L}{\p {\dot \lambda}^\mu}
 =\hf \eta_{\mu \nu} \lambda^\nu~, ~~~~~p_\mu^{({\bar \lambda})} = \frac{\p L}{\p
 {\dot {\bar \lambda}}^\mu} =\hf \eta_{\mu \nu} {\bar \lambda}^\nu .
\lbl{3.47}
\ear

Instead of the coordinates $\lambda^a=(\lambda^\mu,{\bar \lambda}^\mu)$, we can
introduce the new coordinates 
\bear \lambda'^a = (\lambda'^\mu,{\bar
\lambda}'^\mu)~,~~~~~~ &&\lambda'^\mu \equiv \xi^\mu = \frac{1}{\sqrt{2}}
(\lambda^\mu-i {\bar \lambda}^\mu),\nonumber \\ &&{\bar \lambda}'^\mu \equiv
{\bar \xi}^\mu = \frac{1}{\sqrt{2}} (\lambda^\mu+i {\bar \lambda}^\mu),
\lbl{3.48} \ear in which the metric is 
\be g'_{ab} = \gam'_a \cdot \gam'_b =
\begin{pmatrix} 0 & \eta_{\mu \nu}\\ \eta_{\mu \nu} & 0 \end{pmatrix} .
\lbl{3.49} 
\ee 
In the new coordinates we have 
\be {\dot \lambda}^a g_{ab}
\lambda^b = {\dot \lambda}'^a g'_{ab} \lambda'^b = {\dot \xi^\mu} \eta_{\mu \nu}
{\bar \xi}^{\nu} +{\dot {\bar \xi}^\mu} \eta_{\mu \nu} \xi^{\nu} .  
\lbl{3.50}
\ee 
Now the pairs of canonically conjugate variables are $(\xi^\mu,\hf {\bar
\xi}_\mu)$ and $({\bar \xi}^\mu,\hf \xi_\mu)$, whereas in the old coordinates
the pairs were
$(\lambda^\mu,\hf {\lambda}_\mu)$ and $({\bar \lambda}^\mu,\hf {\bar
\lambda}_\mu)$, which was somewhat unfortunate, because the variables in the pair
were essentially the same.

The interaction term can be included by replacing the $\tau$-derivative in the
action (\ref{3.45}) with the covariant derivative:  
\be {\dot Z}^A \rightarrow
{\dot Z}^A + {A^A}_C Z^C .  
\lbl{3.51} 
\ee 
So we obtain\,\ci{PavsicSympl}
\be I = \hf \int \dd \tau \, ({\dot Z}^A + {A^A}_C Z^C)G_{AB} Z^B .  
\lbl{3.52}
\ee
 This is a generalized Bars action\,\ci{Bars}, invariant under
$\tau$-dependent (local) rotations of $Z^A$.  As discussed in
\ci{PavsicSympl}, the gauge fields ${A^A}_C (\tau)$ are not dynamical; they
have the role of Lagrange multipliers, whose choice determines a gauge, related
to the way of how the canonically conjugated variables can be locally rotated
into each other.

For a particular choice of ${A^A}_C$, we obtain 
\be {A^A}_C Z^C G_{AB} Z^B =
  \alpha \, p^\mu p_\mu + \beta \,\lambda^\mu p_\mu 
  + \gam\, {\bar \lambda}^\mu p_\mu .
\lbl{3.53} 
\ee 
Here $\alpha$, $\beta$, $\gam$ are Lagrange multipliers
contained in ${A^A}_C$.  Other choices of ${A^A}_C$ are possible, and they give
expressions that are different from (\ref{3.53}).  A nice theory of how its works
in the bosonic subspace, was elaborated by Bars (see, e.g., refs.\ci{Bars}).

The action (\ref{3.52}), for the case (\ref{3.53}), gives the constraints 
\be
p^\mu p_\mu = 0~,~~~~~ \lambda^\mu p_\mu=0~,~~~~ {\bar \lambda}^\mu p_\mu=0,
\lbl{3.54} 
\ee 
or equivalently 
\be p^\mu p_\mu = 0~,~~~~~ \xi^\mu p_\mu=0~,~~~~
{\bar \xi}^\mu p_\mu=0, 
\lbl{3.54a} 
\ee 
if we use coordinates $\xi^a
=(\xi^\mu,{\bar \xi})$, defined in Eq.\,(\ref{3.48}).

Upon quantization we have 
\be {\hat p}^\mu {\hat p}_\mu \Psi = 0~,~~~~~ {\hat
\lambda}^\mu {\hat p}_\mu \Psi=0 ~,~~~~ {\hat{\bar \lambda}}^\mu {\hat p}_\mu
\Psi=0,
 \lbl{3.55} 
 \ee or equivalently 
\be {\hat p}^\mu {\hat p}_\mu \Psi =
0~,~~~~~ {\hat \xi}^\mu {\hat p}_\mu \Psi=0~, ~~~~ {\hat {\bar \xi}}^\mu {\hat
p}_\mu \Psi=0, 
\lbl{3.55a} 
\ee 
where the quantities with hat are are operators,
satisfying 
\be [{\hat x}^\mu,{\hat p}^\nu] = i \eta^{\mu \nu}~, ~~~~ [{\hat
x}^\mu,{\hat x}^\nu]=0,~~~~[{\hat p}^\mu,{\hat p}^\nu]=0, 
\lbl{3.55b}
\ee 
\be \{{\hat \lambda}^\mu,{\hat \lambda}^\nu \} = 2 i \eta^{\mu \nu}~, 
~~~~\{{\hat {\bar \lambda}}^\mu,{\hat {\bar \lambda}}^\nu \}= 2i \eta^{\mu
\nu}~,~~~~ \{{\hat \lambda}^\mu,  {\hat {\bar \lambda}}^\nu \}  = 0.
\lbl{3.55d} 
\ee
\be
\{{\hat \xi}^\mu,{\hat {\bar \xi}}^\nu\} = \eta^{\mu \nu}~, ~~~~ \{{\hat
\xi}^\mu,{\hat \xi}^\nu\}=0, 
~~~~\{{\hat {\bar \xi}}^\mu,{\hat {\bar \xi}}^\nu\}=0,
\lbl{3.55c} 
\ee 
The
operators can be represented as 
\be {\hat x}^\mu \rightarrow x^\mu~,~~~~{\hat
p}_\mu \rightarrow - i \frac{\p}{\p x^\mu}~,~~~~{\hat \xi}^\mu \rightarrow
\xi^\mu , ~~~~{\hat {\bar \xi}}^\mu \rightarrow \frac{\p}{\p \xi^\mu} ,
\lbl{3.56} 
\ee 
where 
\be x^\mu x^\nu - x^\nu x^\mu = 0~,~~~~~\xi^\mu \xi^\nu +
\xi^\nu \xi^\mu = 0.  
\lbl{3.57} 
\ee 
A state $\Psi$ can be represented as a wave
function $\psi(x^\mu,\xi^\mu)$ of commuting coordinates $x^\mu$ and
anticommuting (Grassmann) coordinates $\xi^\mu$.

In Eq.\,(\ref{3.55}) we have two copies of the Dirac equation, where ${\hat
\lambda}^\mu$ and ${\hat {\bar \lambda}}^\mu$ satisfy the Clifford algebra
anticommutation relations (\ref{3.55d}), and are related to $\gam^\mu$,
${\hat \gam^\mu}$ according to
\be {\hat \lambda}^\mu
= \gam^\mu~, ~~~~~~~{\hat {\bar \lambda}}^\mu = i {\bar \gam}^\mu .  
\lbl{3.60}
\ee 
Using (\ref{3.56}), we find that the quantities $\gam_\mu$, ${\bar
\gam}_\mu$, satisfying 
\be \gam_\mu \cdot \gam_\nu = \eta_{\mu \nu}~,~~~~~~
{\bar \gam}_\mu \cdot {\bar \gam}_\nu = \eta_{\mu \nu} .  
\lbl{3.60a} 
\ee 
can be
represented according to 
\be \gam_\mu = \frac{1}{\sqrt{2}} \left ( \xi_\mu +
\frac{\p}{\p \xi^\mu} \right )~,~~~~~ {\bar \gam}_\mu = \frac{1}{\sqrt{2}} \left
( \xi_\mu - \frac{\p}{\p \xi^\mu} \right ) .  
\lbl{3.61} 
\ee

If we expand $\psi(x^\mu,\xi^\mu)$ in terms of the Grassmann variables
$\xi^\mu$, we obtain a finite number (i.e., $2^n$) of terms:  
\be \psi(x^\mu,\xi^\mu)
= \sum_{r=0}^n \psi_{\mu_1 \mu_2 ...\mu_r} \, \xi^{\mu_1} \xi^{\mu_2}
...\xi^{\mu_r} .  
\lbl{3.63} 
\ee 
In the case of 4D spacetime, $n=4$, the wave
function has $2^4=16$ components.  The state $\Psi$ can then be represented as a
column $\psi^\alpha (x)$, $\alpha = 1,2,..,16$, and the operators $\gam^\mu$,
${\bar \gam}^\mu$ as $16 \times 16$ matrices.  Because we have built our theory over the $8D$
phase space, our spinor has not only four, but sixteen components.  This gives a
lot of room for unified theories of particles and
fields\,\ci{Zenczykowski,Castro8D,PavsicKaluza,PavsicKaluzaLong,PavsicE8}.

\section{Basis vectors, Clifford algebras, spinors and quantized
fields}\lbl{sec4}

\subsection{Spinors as particular Clifford numbers}

We have seen that the generators of Clifford algebras have the properties of
quantum mechanical operators.  Depending on the kind of Clifford algebra, they
satisfy the commutation or anti commutation relations for bosonic or fermionic
creation and annihilation operators.

From the operators $\theta_\mu$ and $\bth_\mu$, defined in Eq.\,(\ref{3.15}), we
can build up spinors by taking a ``vacuum" 
\be \Omega = \prod_{\mu} \bth_\mu~
,~~~\text{which satisfies}~~~~ \bth_\mu \Omega = 0,
\lbl{4.1} 
\ee 
and acting on it by
``creation" operators $\theta_\mu$.  So we obtain a ``Fock space" basis for
spinors\,\ci{Winnberg,BudinichP,PavsicInverse,PavsicMoskva13,BudinichM} 
\be s_\alpha = ({\bf 1} \Omega,\, \theta_\mu \Omega,\, \theta_\mu
\theta_\nu \Omega,\, \theta_\mu \theta_\nu \theta_\rho \Omega,\, \theta_\mu
\theta_\nu \theta_\rho \theta_\sigma \Omega ),
 \lbl{4.3} 
 \ee 
 in terms of which
any state can be expanded as 
\be \Psi_\Omega = \sum \psi^{\alpha} s_\alpha ,~~~~~~~ \alpha = 1,2,...,2^n .
\lbl{4.4} 
\ee
Components $\psi^\alpha$ can be spacetime dependent fields. 
With the operators $\theta_\mu$, $\bth_\mu$ we can construct
spinors as the elements of a minimal left ideal of a Clifford algebra $Cl(2n)$.
We will take the dimension of spacetime $n=4$, so that our phase space will have
dimension 8, and the Clifford algebra, built over it, will be $Cl(2,6)$
which we will simply denote $Cl(8)$ or, in general, $Cl(2n)$.

Besides (\ref{4.1}), there are other possible vacuums, e.g., 
\bear
  \Omega = \prod_\mu \theta_\mu ~,~~~~~~&&\theta_\mu \Omega = 0, 
  \lbl{4.5}\\
\Omega = \left
( \prod_{\mu \in R_1} \theta_\mu \right ) \left (\prod_{\mu \in R_2} \bth_\mu
\right ) , &&\theta_\mu \Omega=0,~~\text{if} ~\mu \in R_2
 \nonumber\\
&&\bth_\mu \Omega=0, ~~\text{if} ~\mu \in R_2.
\lbl{4.6}
\ear
where
\be
   R_1=\{\mu_1,\mu_2,...,\mu_r\}~,~~~~~~~~~R_2=\{\mu_{r+1},\mu_{r+2},...,
  \mu_n\}
\lbl{4.6aa}
\ee
There are $2^n$ vacuums of such a kind.
By taking all those vacuums, we obtain the Fock space basis for the whole
$Cl(2n)$.  If $n=4$, the latter algebra consists of 16 independent minimal left
ideals, each belonging to a different vacuum (\ref{4.6}) and containing
16-component spinors ($2^n=16$ if $n=4$), such as (\ref{4.4}). A generic
element of $Cl(8)$ is the sum of the spinors $\Psi_{\Omega_i}$,
$i=1,2,3,4,...,16$. belonging to the ideal associated with a vacuum
$\Psi_{\Omega_i}$:
\be
  \Psi = \sum_i \Psi_{\Omega_i} = \psi^{\alpha i} s_{\alpha i}
  \equiv \psi^{\tl A} s_{\tl A}~,~~~~~~~~
  {\tl A} = 1,2,3,4,...,256 ,
\lbl{4.6a}
\ee
where $s_{\tl A}\equiv s_{\alpha i}$, $\alpha,i=1,2,...,16$, is the Fock space basis for $Cl(8)$,
and $\psi^{\tl A}\equiv \psi^{\alpha i}$ are spacetime dependent fields.
The same element $\Psi \in Cl(8)$ can be as well expanded in terms of
the multivector basis,
\be
  \Psi = \psi^A \gam_A~,~~~A=1,2,3,4,...,256,
\lbl{4.6b}
\ee
where
\be
  \gam_A = {\ul 1},\,\gam_{a_1},\,\gam_{a_1} \wg \gam_{a_2},...,
  \gam_{a_1} \wg \gam_{a_2}\wg ...\wg \gam_{a_{2 n}},
\lbl{4.6c}
\ee
which can be written compactly as
\be
  \gam_A = \gam_{a_1} \wg \gam_{a_2} \wg ...\wg \gam_{a_r}~,~~~~~~
  r=0,1,2,...,2 n .
\lbl{4.6d}
\ee

We see that if we construct the Clifford algebra of the 8-dimensional phase
space, then we have much more room for unification\footnote{
Unification based on Clifford algebras in phase space was considered
by Zenczykowski\,\ci{Zenczykowski}}
 of elementary particles and
fields than in the case of $Cl(1,3)$, constructed over $4D$ spacetime.  We have
a state $\Psi$ that can be represented by a $16 \times 16$ matrix, whose elements
can represent all known particles of the 1$^{\rm st}$ generation of the Standard
model. Thus, 64 elements of this $16 \times 16$ matrix include the left
and right handed (L,R)
versions of the states $(e,\nu_e)$, $(u,d)_r$, $(u,d)_b$, $(u,d)_g$, and their
antiparticles, times factor two, because all those states, satisfying
the generalized Dirac equation\ci{PavsicKaluza,PavsicKaluzaLong}(see also
Sec.\ref{sec5}.2 ), can in principle be
superposed with complex amplitudes.

If we take {\it space inversion} (P) of those 64 states by using the same
procedure as in Ref.\,\ci{PavsicInverse}, we obtain another 64
states of the $16 \times 16$ matrix representing $\Psi$, namely the states
of mirror particles (P-particles).  Under time
reversal (T) (see Ref.\,\ci{PavsicInverse}), we obtain yet another 64 states corresponding to time reversed
particles (T-particles).  And finally, under PT, we obtain 64 states of time
reversed mirror particles (PT-particles).  Altogether, we have $4 \times 64 =
256$ states:  

\be
  s_{\alpha i} = \begin{pmatrix}
\begin{pmatrix} e & \nu\\ {\bar e} & {\bar \nu} \end{pmatrix} & \begin{pmatrix}
e & \nu\\ {\bar e} & {\bar \nu} \end{pmatrix}_P & \begin{pmatrix} e & \nu\\
{\bar e} & {\bar \nu} \end{pmatrix}_T & \begin{pmatrix} e & \nu\\ {\bar e} &
{\bar \nu} \end{pmatrix}_{PT}\\ \begin{pmatrix} u & d\\ {\bar u} & {\bar d}
\end{pmatrix}_r & \begin{pmatrix} u & d\\ {\bar u} & {\bar d}
\end{pmatrix}_{r,P} & \begin{pmatrix} u & d\\ {\bar u} & {\bar d}
\end{pmatrix}_{r,T} & \begin{pmatrix} u & d\\ {\bar u} & {\bar d}
\end{pmatrix}_{r,PT}\\ \begin{pmatrix} u & d\\ {\bar u} & {\bar d}
\end{pmatrix}_g & \begin{pmatrix} u & d\\ {\bar u} & {\bar d}
\end{pmatrix}_{g,P} & \begin{pmatrix} u & d\\ {\bar u} & {\bar d}
\end{pmatrix}_{g,T} & \begin{pmatrix} u & d\\ {\bar u} & {\bar d}
\end{pmatrix}_{g,PT}\\ \begin{pmatrix} u & d\\ {\bar u} & {\bar d}
\end{pmatrix}_b & \begin{pmatrix} u & d\\ {\bar u} & {\bar d}
\end{pmatrix}_{b,P} & \begin{pmatrix} u & d\\ {\bar u} & {\bar d}
\end{pmatrix}_{b,T} & \begin{pmatrix} u & d\\ {\bar u} & {\bar d}
\end{pmatrix}_{b,PT} \end{pmatrix} , \lbl{4.8} \ee
where
\be 
  \begin{pmatrix} e & \nu\\ {\bar e} & {\bar \nu} \end{pmatrix} 
   \equiv\begin{pmatrix} e_L & i
e_L & \nu_L & i \nu_L\\ e_R & i e_R & \nu_R & i \nu_R\\ {\bar e}_L & i {\bar
e}_L & {\bar \nu}_L & i {\bar \nu_L}\\ {\bar e}_R & i {\bar e}_R & {\bar \nu}_R
& i {\bar \nu}_R \end{pmatrix} , 
\lbl{4.7} 
\ee 
and similarly for $u,~d$.

Those states interact with the corresponding gauge fields\footnote{
How this works in the case of $Cl(1,3)$ is shown in Ref.\,\ci{PavsicInverse}
(see also \ci{PavsicKaluza,PavsicKaluzaLong})},
which include the
gauge fields of the Standard model, such as the photon, weak bosons and gluons.
There exist also mirror versions, as well as T and PT versions of the standard
gauge bosons\footnote{ Here we extend the concept of mirror particles and mirror
gauge fields. The idea of mirror particles was first put forward
by Lee and Yang\,\ci{LeeYang}
who realized that ``...there could exist corresponding elementary particles exhibiting
opposite asymmetry such that in the broader sense there will still be over-all
right-left symmetry." Further they wrote: ``If this is the case, it should be
pointed out that there must exist two kinds of protons $p_R$ and $p_L$, the right-handed
one and the left-handed one." Lee and Yang thus considered the possibility of
mirror particles, though they did not name them so, and as an example they considered
ordinary and mirror protons. Later, Kobzarev et al.\,\ci{Okun}, instead
of P-partners, considered CP-partners of ordinary particles and called them
``mirror particles".
They argued that a complete doubling of the known particles and forces, except
gravity, was necessary. Subsequently, the idea of mirror particles has been
pursued in refs\,\ci{PavsicMirror}--\ci{FootVolkas}. The connection between
mirror particles and dark matter was suggested in Ref.\,\,\ci{Blinnikov},
and later  explored in many works,
e.g., in \ci{Hodges}--\ci{CiarcellutiFoot}. An explanation of mirror
particles in terms of algebraic spinors (elements of Clifford algebras)
was exposed in Refs.\,\ci{PavsicInverse,PavsicMoskva13}. For a recent review
see \ci{FootReview14}.}.

Of the 256 particle states in Eq.\,(\ref{4.8}), only 1/4 interact with our usual
photons, whereas the remaining 3/4 do not interact with our photons, but they
may interact with mirror photons, T-photons or PT-photons.  This scheme thus
predicts the existence of {\it dark matter}. If the matter in the universe
were evenly distributed over the ordinary particles, P-particles, T-particles
and PT-particles, then $1/4$  of the matter would be visible, and 3/4 dark.
In reality, the distribution of matter in the universe need not be even over
the four different version of the particles. It can deviate from even
distribution, but we expect that the deviation is not very big.
According to the current
astronomical observations about 81.7\% of matter in the universe is dark, and
only 18.3\% is visible.  This roughly corresponds to the ratio 1/4 of the ``visible
states" in matrix (\ref{4.8}).

\subsection{Quantized fields as generalized Clifford numbers}

We can consider a field as an infinite dimensional vector.  As an example, let us
take 
\be \Psi = \psi^{i(x)} h_{i(x)} \equiv \int \dd^n x \,\psi^i (x) h_i (x) ,
\lbl{4.9} 
\ee 
where $i =1,2$, $x\in {\mathbb R}^3$ or $x\in {\mathbb R}^{1,3}$
are, respectively, a discrete index, and $(x)$ a continuous index, denoting,
e.g., a point in 3D space, or an event in 4D spacetime.  The infinite
dimensional vector $\Psi$ is decomposed with respect to an infinite dimensional
basis, consisting of vectors $h_{i(x)} \equiv h_i (x)$, satisfying\,\ci{PavsicSympl} 
\be h_{i(x)}
\cdot h_{j(x')} \equiv \text{$\frac{1}{2}$} (h_{i(x)} h_{j(x')} + h_{j(x')}
h_{i(x)}) = \rho_{i(x)j(x')} , 
\lbl{4.10} 
\ee 
where $\rho_{i(x)j(x')}$ is the
metric of the infinite dimensional space ${\cal S}$.  The latter space may in
general have non vanishing curvature\,\ci{PavsicBook}.  If, in particular, the
curvature of ${\cal S}$ is ``flat", then we may consider a parametrization of
${\cal S}$ such that 
\be \rho_{i(x)j(x')} = \delta_{ij} \delta (x-x') .
\lbl{4.11} 
\ee 
In Eq.\,(\ref{4.10}) we have a generalization of the Clifford
algebra relations (\ref{2.4}) to infinite dimensions.

Instead of the basis in which the basis vectors satisfy Eq.\,(\ref{4.10}), we
can introduce the Witt basis 
\bear &&h_{(x)} = \text{$\frac{1}{\sqrt{2}}$}
  (h_{1(x)} + i h_{2(x)}), \lbl{4.12}\\
  &&\bh_{(x)} = \text{$\frac{1}{\sqrt{2}}$}
  (h_{1(x)} - i h_{2(x)}), \lbl{4.13} \ear
in which we have 
\be
  h_{(x)} \cdot \bh_{(x')} = \delta_{(x)(x')}, 
\lbl{4.14} 
\ee 
\be h_{(x)} \cdot h_{(x')} = 0~,~~~~~~\bh_{(x)} \cdot \bh_{(x')} = 0.  
\lbl{4.15} 
\ee 
The vector $h_{i(x)}$
and the corresponding components $\psi^{i(x)}$ may contain an implicit discrete
index $\mu=0,1,2,...,n$, so that Eq.\,(\ref{4.9}) explicitly reads 
\be \Psi =
\psi^{i \mu (x)} h_{i \mu (x)} = \psi^{\mu (x)} h_{\mu (x)} + {\bar \psi}^{\mu
(x)} \bh_{\mu (x)}.  
\lbl{4.16} 
\ee 
Then, Eqs.\, (\ref{4.14}),(\ref{4.15}) become the anticommuting relations
for fermion fields:
\be
   h_{\mu (x)} \cdot \bh_{\nu (x')} = \eta_{\mu \nu} \delta_{(x)(x')},
\lbl{4.17} 
\ee 
\be
  h_{\mu (x)} \cdot h_{\nu (x')}=0~,~~~~~, \bh_{\mu (x)} \cdot\bh_{\nu (x')}=0.  
\lbl{4.18} 
\ee 
The quantities $h_{\mu (x)}$,$ \bh_{\mu (x)}$
are a generalization to infinite dimensions of the Witt basis vectors
$\nth_\mu$, $\bth_\mu$, defined in Eq.\,(\ref{3.15}).

Using $\bh_{\mu(x)}$, we can define a vacuum state as the product\,\ci{PavsicSympl}
\be \Omega =
\prod_{\mu,x} \bh_{\mu (x)} ~, ~~~~~~~\bh_{\mu (x)} \Omega = 0.  
\lbl{4.19} 
\ee
Then, using the definition (\ref{4.16}) of a vector $\Psi$, we have 
\be \Psi
\Omega = \psi^{\mu (x)} h_{\mu (x)} \Omega .  
\lbl{4.20} 
\ee 
Because
$\bh_{\mu(x)} \Omega = 0$, the second part of $\Psi$ disappears in the above
equation.

The infinite dimensional vector $\Psi$, defined in Eq.\,(\ref{4.16}), consists
of two parts, $\psi^{\mu(x)} h_{\mu(x)}$ and ${\bar \psi}^{\mu(x)}
\bh_{\mu(x)}$, which both together span the phase space of a field theory.

The vector $\psi^{\mu(x)} h_{\mu(x)}$ can be generalized to an element of an
infinite dimensional Clifford algebra:  
\be \psi_0 {\ul1 }+\psi^{\mu (x)} h_{\mu(x)}
+\psi^{\mu(x)\nu(x')} h_{\mu(x)}h_{\nu(x')} + ...  
\lbl{4.22}
 \ee 
 Acting with
the latter object on the vacuum (\ref{4.19}), we obtain 
\be \Psi_\Omega =
(\psi_0 {\ul 1} +\psi^{\mu (x)} h_{\mu(x)} +\psi^{\mu(x)\nu(x')} h_{\mu(x)}h_{\nu(x')} +
...)  \Omega .  
\lbl{4.23} 
\ee 
This state is the infinite dimensional space
analog of the spinor as an element of a left ideal of a Clifford algebra.  At a
fixed point $x\equiv x^\mu$ there is no ``sum" (i.e., integral) over $x$ in
expression (\ref{4.22}), and we obtain a spinor with $2^{n}$ components.  It
is an element of a minimal left ideal of $Cl(2 n)$.  In 4D spacetime, $n=4$, and
we have $Cl(8)$ at fixed $x$.

Besides the vacuum (\ref{4.19}) there are other vacuums, such as 
\be \Omega =
\prod_{\mu,x} h_{\mu(x)} ~, ~~~~~~h_{\mu(x)} \Omega = 0 , 
\lbl{4.24} 
\ee 
and, in general, 
\be \Omega = \left ( \prod_{\mu \in R_1,x} \bh_{\mu(x)} \right ) \left
( \prod_{\mu \in R_2,x} h_{\mu(x)} \right ) .  
\lbl{4.25} 
\ee
 Here $R=R_1 \cup
R_2$ is the set of indices $\mu=0,1,2,...,n$, and $R_1$, $R_2$ are subsets of
indices, e.g., $R_1=\{1,3,5,...,n\}$, $R_2 = \{2,4,...,n-1\}$.

Expression (\ref{4.25}) can be written as 
\be \Omega = \prod_x \left ( \prod_{\mu
\in R_1} \bh_{\mu(x)} \right ) \left ( \prod_{\mu \in R_2} h_{\mu(x)} \right ) =
\prod_x \Omega_{(x)} , 
\lbl{4.26} 
\ee 
where 
\be \Omega_{(x)} = \left ( \prod_{\mu
\in R_1} \bh_{\mu(x)} \right ) \left ( \prod_{\mu \in R_2} h_{\mu(x)} \right ) ,
\lbl{4.27} 
\ee 
is a vacuum at a fixed point $x$.  At a fixed $x$, we have
$2^{n}$ different vacuums, and thus $2^{n}$ different spinors, defined
analogously to the spinor (\ref{4.23}), belonging to different minimal ideal of
$Cl(2n)$.

The vacuum (\ref{4.25}) can be even further generalized by taking different
domains ${\cal R}_1$, ${\cal R}_2$ of spacetime positions $x$:  
\be \Omega =
\left ( \prod_{\mu \in R_1,x\in {\cal R}_1} \bh_{\mu(x)} \right ) \left (
\prod_{\mu \in R_2,x\in {\cal R}_2} h_{\mu(x)} \right ) , 
\lbl{4.28} 
\ee 
In such
a way we obtain many other vacuums, depending on a partition of ${\mathbb
R}^n$ into two domains ${\cal R}_1$ and ${\cal R}_2$ so that ${\mathbb R}^n =
{\cal R}_1 \cup {\cal R}_2$.

Instead of the configuration space, we can take the momentum space, and
consider, e.g., positive and negative momenta.  In Minkowski spacetime we can
have a vacuum of the form
\be \Omega = \left ( \prod_{\mu,  p^0>0,{\bf p}}
\bh_{\mu(p^0,{\bf p})} \right ) \left ( \prod_{\mu,p^0<0,{\bf p}}
h_{\mu(p^0,\bf{p})} \right ) ,
\lbl{4.29} 
\ee
which is annihilated according to
\be \bh_{\mu(p^0>0,{\bf p})} \Omega = 0 ~,~~~~~~ h_{\mu(p^0<0,{\bf p})} \Omega
=0.  
\lbl{4.30} 
\ee

For the vacuum (\ref{4.29}), $\bh_{\mu(p^0>0,{\bf p})}$ and $h_{\mu(p^0<0,{\bf
p})}$ are {\it annihilation operators}, whereas $\bh_{\mu(p^0<0,{\bf p})}$ and
$h_{\mu(p^0>0,{\bf p})}$ are {\it creation operators} from which one can compose
the states such as
$$
   \Bigl(\psi_0 {\ul 1} + \psi^{\mu(p^0>0,{\bf p})} h_{\mu(p^0>0,{\bf p})}
   +\psi^{\mu(p^0>0,{\bf p})} \psi^{\nu(p'^0 >0,{\bf p}')} h_{\mu(p^0>0,{\bf p})} 
    h_{\nu(p'^0>0,{\bf p}')}   + ...$$
\be
   \hs{2cm} ... \,+\psi^{\mu(p^0<0,{\bf p})} \bh_{\mu(p^0<0,{\bf p})} + ...
   \Bigr) \Omega .
\lbl{4.31}
\ee

The vacuum, satisfying (\ref{4.30}), has the property of the {\it bare} Dirac
vacuum. This can be seen if one changes the notation according to
\bear
  && h_\mu (p^0>0,{\bf p}) \equiv b_\mu^\dg ({\bf p})~,~~~~
  \bh_\mu (p^0>0,{\bf p}) \equiv b_\mu ({\bf p}) \lbl{4.32}\\
  &&{\bh}_\mu (p^0<0,{\bf p}) \equiv d_\mu ({\bf p})~,~~~~
  h_\mu (p^0<0,{\bf p}) \equiv d_\mu^\dg ({\bf p}) \lbl{4.33}\\
  && \Omega \equiv |0 \rangle_{\rm bare} \lbl{4.33a}.
\ear
A difference with the usual Dirac theory is that our operators have index $\mu$
which takes four values, and not only two values, but otherwise the principle
is the same.

The operators $b_\mu^\dg$ and $b_\mu$, respectively, create and annihilate
a positive energy fermion, whereas the operators $d_\mu$,  $d_\mu^\dg$
create and annihilate a negative energy fermion. This is precisely a property
of the bare Dirac vacuum. Instead of the bare vacuum, in quantum field
theories we consider the the physical vacuum
\be
  |0 \rangle = \prod_{\mu,{\bf p}} d_\mu ({\bf p}) |0 \rangle_{\rm bare} ,
\lbl{4.34}
\ee
in which the negative energy states are filled, and which in our notation
reads
\be
  \Omega_{\rm phys} = \prod_{\mu,p^0<0,{\bf p}} \bh_{\mu(p^0,{\bf p})} \Omega .
\lbl{4.35}
\ee

We see that in a field theory \`a la Clifford, a vacuum is defined as the
product of fermionic operators (generators in the Witt basis). The Dirac
(physical) vacuum is defined as a sea of negative energy states according
to (\ref{4.34}) or (\ref{4.35}). Today it is often stated that the Dirac
vacuum as the sea of negative energy states is an obsolete concept. But
within a field theory based infinite dimensional Clifford algebras,
a vacuum is in fact a ``sea" of states defined by infinite (uncountable)
product of operators.

With respect to the vacuum (\ref{4.29}), one kind of particles are created
by the positive energy operators $h_\mu (p^0>0,{\bf p})$, whilst the other kind
of particles are created by the negative energy operator,
${\bar h}_\mu (p^0<0,{\bf p})$. The vacuum with reversed properties can also
be defined, besides many other possible vacuums. All those vacuums participate
in a description of the interactive processes of elementary particles.
What we take into account in our current quantum field theory calculations
seem to be only a part of a larger theory that has been neglected. It could be
that some of the difficulties (e.g., infinities) that we have encountered in
QFTs so far, are partly due to neglect of such a larger theory.

In an analogous way we can also construct\,\ci{PavsicSympl} bosonic states
as elements of an infinite dimensional symplectic Clifford algebra.
The generators of the latter algebra are bosonic field operators. We will use
them in the next subsection when constructing the action and field equations.

\subsection{The action and field equations}

A sympletic vector is\,\ci{PavsicSympl}
\bear
  \Phi &=& \phi^{i(x)}k_{i(x)} = \phi^{1(x)}k_{1(x)} +\phi^{2(x)}k_{2(x)}\nonumber \\
  &\equiv& \phi^{(x)} k_{(x)}^\phi + \Pi_{(x)} k_{(x)}^\Pi , \nonumber\\
  &&x \in {\mathbb R}^3~~~{\rm or}~~~x \in {\mathbb R}^{1,3}.
\lbl{4.36}
\ear
Here $\phi^{i(x)} = (\phi^{(x)},\Pi^{(x)})$ are components and $k_{i(x)}$, $i=1,2$,
basis vectors, satisfying
\be
  k_{i(x)} \wg k_{j(x')} = \frac{1}{2}[k_{i(x)}, k_{j(x')}] = J_{i(x) j(x')},
\lbl{4.37}
\ee
where
\be
  J_{i(x) j(x')}= \begin{pmatrix}
                         0  &  \delta_{(x)(x')}\\
                        -\delta_{(x)(x')} & 0
                         \end{pmatrix}
\lbl{4.37a}
\ee

The action is
\be
  I = \int \dd \tau \, \left [ \hf {\dot \phi}^{i(x)} 
  J_{i(x) j(x')} \phi^{j(x')} - H \right ],
\lbl{4.38}
\ee
where
\be
  H = \hf \phi^{i(x)} K_{i(x)j(x')} \phi^{j(x')}
\lbl{4.39}
\ee
is the Hamiltonian, and
\be
  \hf {\dot \phi}^{i(x)} J_{i(x) j(x')} \phi^{j(x')}
   = \hf (\Pi {\dot \phi} - \phi {\dot \Pi})
\lbl{4.40}
\ee
the symplectic form.

In particular\,\ci{PavsicSympl}, if $x\equiv x^r \in {\mathbb R}^3$, $r=1,2,3$,
and
\be
  K_{i(x)j(x')} = \begin{pmatrix}
       (m^2 + \p^r \p_r ) \delta (x-x')  &  0 \\
         0     &   \delta (x-x')
         \end{pmatrix},
\lbl{4.41}
\ee
then we obtain the phase space action for a classical scalar field.

If $x \equiv x^r$, $r=1,2,3$, and
\be
  K_{i(x)j(x')} = \left ( - \frac{1}{2 m}\p^r \p_r + V(x) \right  ) \delta(x-x') 
  g_{ij}~,~~~~~~
  g_{ij} = \begin{pmatrix} 0 & 1\\ 1 & 0 \end{pmatrix},
\lbl{4.42}
\ee
then the action (\ref{4.38}) describes the classical Schr\"odinger field.

If $x \equiv x^\mu \in {\mathbb R}^{1,3}$, $\mu=0,1,2,3$, and
\be
  K_{i(x)j(x')} = \left ( - \frac{1}{2 \Lambda}\p^\mu \p_\mu \right  ) 
  \delta(x-x')  g_{ij}~,~~~~~~
  g_{ij} = \begin{pmatrix} 0 & 1\\ 1 & 0 \end{pmatrix},
\lbl{4.43}
\ee
then from (\ref{4.38}) we obtain the action for the classical Stueckelberg field.

From the action (\ref{4.38}) we obtain the following equations of motion
\be
  {\dot \phi}^{i(x)} = J^{i(x)j(x')} \frac{\p H}{\p \phi^{j(x')}},
\lbl{4.44}
\ee
where $\p/\p \phi^{j(x')} \equiv \delta / \delta \phi^{j(x')}$ is the functional
derivative. By following the analogous procedure as in Sec.\,\ref{sec3}.2,
we obtain,\ci{PavsicSympl} the equations of motion for the operators:
\be
  {\dot k}_{j(x')} = k^{i(x)} K_{i(x)j(x')}= [k_{j(x')}, {\hat H}] ,
\lbl{4.45}
\ee
where
\be
  {\hat H} = \hf k^{i(x)} K_{i(x)j(x')} k^{j(x')}.
\lbl{4.46}
\ee

The Heisenberg equations of motion (\ref{4.45}) can be derived from the
action
\be
  I = \hf \int \dd \tau \Bigl( {\dot k}^{i(x)} J_{i(x)j(x')} k^{j(x')} +
  k^{i(x)} K_{i(x)j(x')} k^{j(x')} \Bigr).
\lbl{4.47}
\ee

The Poisson bracket between two functionals of the classical phase space
fields is
\be
  \{ f(\phi^{i(x)}),g(\phi^{j(x')}) \}_{PB} = \frac{\p f}{\p \phi^{i(x)}}
  J^{i(x) j(x')} \frac{\p g}{\p \phi^{j(x')}} .
\lbl{4.48}
\ee

In particular, if $f= \phi^{k(x'')}$, $g=\phi^{\ell (x''')}$, Eq.\,(\ref{4.48})
gives\,\ci{PavsicSympl}
\be
  \{\phi^{k(x'')},\phi^{\ell (x''')} \}_{PB} = J^{k(x'') \ell (x''')}
  = k^{k(x'')} \wg k^{\ell (x''')} \equiv \hf [k^{k(x'')},k^{\ell (x''')} ].
\lbl{4.49}
\ee
On the one hand, the Poisson bracket of two classical fields is equal to the
symplectic metric. On the other hand, the symplectic metric is equal
to the wedge product of basis vectors. In  fact, the basis vectors are
quantum mechanical operators, and  they satisfy the quantum mechanical
commutation relations
\be
  \hf [ k_\phi (x),k_\Pi (x')] = \delta (x-x'),
\lbl{4.50}
\ee
or
\be
  [{\hat \phi} (x), {\hat \Pi} (x')] = i \delta (x-x'),
\lbl{4.51}
\ee
if we identify $\frac{1}{{\sqrt 2}} k_\phi (x) \equiv {\hat \phi} (x)$,
$\frac{i}{{\sqrt 2}} k_\Pi (x') \equiv {\hat \Pi} (x')$

A similar procedure can be repeated for fermionic vectors\,\ci{PavsicSympl}.

\section{Towards quantum gravity}\lbl{sec5}

\subsection{Gravitational field from Clifford algebra}

The generators of a Clifford algebra, $\gam_\mu$, ${\bar \gam}_\mu$, are
 (i) tangent vectors to a manifold which, in particular, can be spacetime. On the
other hand, (ii) the $\gam_\mu$, ${\bar \gam}_\mu$ are superpositions of 
fermionic creation and annihilation operators, as shown in Eqs.\,(\ref{3.15a}),
(\ref{3.16}). This two facts, (i) and (ii), must have profound and far reaching
consequences for quantum gravity. Here I am going to expose some further ingredients
that in the future, after having been fully investigated, will illuminate the
relation between quantum theory and gravity.

As a first step let us consider a generalized spinor field, defined
in Sec.\,\ref{sec4}.1:
\be
  \Psi = \psi^{\tl A} s_{\tl A} = \phi^A \gam_A .
\lbl{5.1}
\ee
We are interested in the expectation value of a a vector $\gam_\mu$ with
respect to the state $\Psi$:
\be
  \langle \gam_\mu \rangle_1 \equiv \langle \Psi^\ddg \gam_\mu \Psi \rangle_1
  = \langle \psi^{* \tl A} s_{\tl A}^\ddg \gam_\mu s_{\tl B} \psi^{\tl B} \rangle_1
\lbl{5.2}
\ee
The subscript 1 means vector part of the expression. Recall from
Sec.\,\ref{sec2} that $\ddg$ means reversion. Taking
\be
  \langle s_{\tl A}^\ddg \gam_\mu s_{\tl B} \rangle_1 =
  C_{\tl A \tl B \mu}^{~c} \gam_c ~,
\lbl{5.3}
\ee
we have
\be
  \langle \gam_\mu \rangle_1 = {e_\mu}^c \gam_c ,
\lbl{5.3a}
\ee
where
\be
  {e_\mu}^c = \psi^{* \tl A} C_{\tl A \tl B \mu}^{~c} \psi^{\tl B}
\lbl{5.4}
\ee
is the fierbein. The vector $\gam_\mu$ gives the flat spacetime metric
\be
  \gam_\mu \cdot \gam_\nu = \eta_{\mu \nu} .
\lbl{5.5}
\ee
The expectation value vector $\langle \gam_\mu \rangle_1$ gives a curved
spacetime metric
\be
  g_{\mu \nu} = \langle \gam_\mu \rangle_1 \cdot \langle \gam_\nu \rangle_1
  = {e_\mu}^c {e_\nu}^d \eta_{cd}
\lbl{5.6}
\ee
which, in general, differs from $\eta_{\mu \nu}$. If $\psi^{\tl A}$ depends
on position $x\equiv x^\mu$ in spacetime, then also ${e_\mu}^c$ depends
on $x$, and so does $g_{\mu \nu}$.

From Eq.\,(\ref{5.3}) we obtain
\be
  {e_\mu}^a = \langle \gam_\mu \rangle \cdot \gam^a .
\lbl{5.7}
\ee
From the fierbein we can calculate the spin connection,
\be
  {\omega_\mu}^{ab} = \hf (e^{\rho b} e_{[\mu,\rho]}^{~a} - 
  e^{\rho a} e_{[\mu,\rho]}^{~b} + e^{\rho b} e^{a \sigma} e_{\mu c} 
  e_{[\sigma,\rho]}^{~c}) .
\lbl{5.7a}
\ee
The curvature is
\be
  {R_{\mu \nu}}^{ab} = \p_\mu {\omega_\nu}^{ab} - \p_\nu {\omega_\mu}^{ab}
  + {\omega_\mu}^{ac} {\omega_{\nu c}}^b - {\omega_\nu}^{ac} {\omega_{\mu c}}^b
\lbl{5.8}
\ee
In order to see whether the curvature vanishes or not, let us calculate
$\omega_{[\mu,\nu]}^{~ab}$ by using (\ref{5.4}) in which  we write
\be
  \psi^{* \tl A} \psi^{\tl B} \equiv \psi^{{\tl A}{\tl B}} .
\lbl{5.9}
\ee
We obtain
\bear
  \omega_{[\mu,\nu]}^{~ab} &=& \hf \left [ C_{\tl A \tl B}^{~b \rho}
  C_{{\tl C}{\tl D} \mu}^a (\psi^{\tl A \tl B} {\psi^{{\tl C}{\tl D}}}_{,\rho} )_{\,\nu}
  -C_{\tl A \tl B}^{~b \rho}
  C_{{\tl C}{\tl D} \rho}^a (\psi^{\tl A \tl B} {\psi^{{\tl C}{\tl D}}}{,_\mu} )_{,\nu}
  \right . \nonumber \\
 && -C_{\tl A \tl B}^{~a \rho}
  C_{{\tl C}{\tl D} \mu}^b (\psi^{\tl A \tl B} {\psi^{{\tl C}{\tl D}}}_{,\rho} )_{,\nu}
  +C_{\tl A \tl B}^{~a \rho}
  C_{{\tl C}{\tl D} \rho}^b (\psi^{\tl A \tl B} {\psi^{{\tl C}{\tl D}}}_{,\mu} )_{,\nu}
   \nonumber \\
  &&\hs{1cm} \left . + \,\text{more terms}\, - \,(\mu \rightarrow \nu, \nu \rightarrow \mu) \right ] .
\lbl{5.10}
\ear
The latter expression does not vanish identically. In general it could be
different from zero, which would mean that also the curvature (\ref{5.8})
is different from zero, and that the generalized spinor field $\psi^{\tl A}(x)$
induces gravitation. This assertion should be checked by explicit calculations
with explicit structure constants $C_{{\tl A}{\tl B}}^{~b \rho}$ and/or
their symmetry relations.

If $\psi^{\tl A} (x)$ indeed induces gravitation, then we have essentially
arrived at the basis of quantum gravity. At the basic level, gravity is thus
caused by a spacetime dependent (generalized) spinor field $\psi^{\tl A} (x)$
entering the expression (\ref{5.4}) for vierbein. If $\psi^{\tl A} (x)$ is
constant, or proportional to ${\rm e}^{i p_\mu x^\mu}$ which, roughly speaking,
means that there is no non trivial matter, then ${R_{\mu \nu}}^{ab}
=0$. This has its counterpart in the (classical)
Einstein's equation which say that matter curves spacetime.

\subsection{Action principle for the Clifford algebra valued field}

Let us assume that the field (\ref{5.1}) satisfies the action
principle\footnote{If reduced to a subspace of the Clifford space,
this action contains a mass term.}
\be
  I = \hf \int \dd^4 x \, \p_\mu \phi^A \p_\nu \phi^B \eta_{AB} \eta^{\mu \nu}
\lbl{5.11}
\ee
for a system of scalar fields $\phi^A$ that may contain an implicit
index $i=1,2$, denoting real and imaginary components. Here $\eta_{AB}$ is
the metric of the 16D Clifford space, whereas $\eta_{\mu \nu}$ is the metric of
the 4D Minkowski space. The action (\ref{5.11}) is not invariant under
reparametrizations of coordinates $x^\mu$ (i.e., of general coordinate
transformations). A possible way to make the action invariant is to
replace $\eta^{\mu \nu}$ with $g^{\mu \nu}$, and include a kinetic term for
$g^{\mu \nu}$. Another possible way is to consider the action
\be
  I = \int \dd^4 x \, {\rm det} (\p_\mu \phi^A \p_\nu \phi^B \eta_{AB})^{1/2} .
\lbl{5.12}
\ee
This is an action for a 4D surface $V_4$, embedded in the 16D space, the
embedding functions being $\phi^A (x^\mu)$. The induced metric on $V_4$ is
\be
  g_{\mu \nu} = \p_\mu \phi^A \p_\nu \phi^B \eta_{AB}
\lbl{5.13}
\ee

The theory based on the nonlinear action (\ref{5.12}) is complicated and
difficult to quantize. Therefore we will return to the action (\ref{5.11}) and
try to explore how far can we arrive in inducing non trivial spacetime
metric according to the lines indicated in Sec.\,\ref{sec5}.1. The equations of
motion derived from (\ref{5.11}) are
\be
   \eta^{\mu \nu} \p_\mu \p_\nu \phi^A = 0.
\lbl{5.14a}
\ee
A field $\phi^A$ that satisfied the latter equation satisfies also the Dirac
like equation
\be
   \gam^\mu \p_\mu \phi^A = 0~,~~~~~~ \gam^\mu \cdot \gam^\nu = \eta^{\mu \nu}.
\lbl{5.14b}
\ee
This is so because of the relation (\ref{5.1}) and the fact that $\psi^{\tl A}$
are spinor components belonging to all left minimal ideals of the considered
Clifford algebra.
Eq.\,(\ref{5.14b}) can be contracted by $\gam_A$, and we obtain the
Dirac-K\"ahler equation
\be
  \gam^\mu \p_\mu \phi^A \gam_A = 0,
\lbl{5.14c}
\ee
where
\be
  \gam_A = (1,\gam_{a_1},\gam_{a_1} \wg \gam_{a_2},..., \gam_{a_1} \wg ...
  \wg \gam_{a_4}) .
\lbl{5.15}
\ee
In (\ref{5.14c}) we have a geometric form of the equation.
We can put it in a sandwich between $\gam^B$ and $\gam_A$, or equivalently,
between $s^{\tl B}$ and $s_{\tl A}$, according to
\be
  \langle \gam^B \gam^\mu \gam_A \rangle_S \p_\mu \phi^A = 0~,~~~~
  \text{or} ~~~~\langle s^{\tl B} \gam^\mu s_{\tl B} \rangle_S 
  \p_\mu \phi^{\tl A} = 0.
\lbl{5.15a}
\ee
Here ``$S$'' denotes scalar part, $\langle ~~ \rangle_0$ multiplied by the
dimension of the spinor space.
Here $\langle \gam^B \gam^\mu \gam_A \rangle_S \equiv {(\gam_\mu)^B}_A$
and $\langle s^{\tl B} \gam^\mu s_{\tl A} \rangle_S \equiv 
{(\gam_\mu)^{\tl B}}_{\tl A}$ are $16 \times 16$ matrices, representing
the vectors $\gam_\mu$. Those matrices are reducible to four $4 \times 4$
blocks
\be
  \langle s^{\beta} \gam^\mu s_{\tl B} \rangle_S \equiv 
{(\gam_\mu)^{\beta}}_{\alpha} ~,~~~~~~\alpha, \beta = 1,2,3,4~~~ \text{spinor
index},
\lbl{5.15b}
\ee
which are just the (usual) Dirac matrices.

Eq.\.(\ref{5.15a}) can be derived from the action
\be
  I = \int \dd^4 x\, \langle \phi^A \gam_A \gam^\mu \gam_B \p_\mu \phi^B
  \rangle_S ,
\lbl{5.16}
\ee
which can also be written in terms of the generalized spinors
$\psi^{\tl A} s_{\tl A}$:
\be
  I = \int \dd^4 x\, \langle \psi^{\tl A} s_{\tl A} \gam^\mu 
  s_{\tl B} \p_\mu \psi^B \rangle_S .
\lbl{5.17}
\ee

The action (\ref{5.16}) or (\ref{5.17}) is not invariant under general coordinate
transformations of $x^\mu$. For this aim one has to consider position dependent
Clifford numbers, giving the connection according to\,\ci{PavsicKaluzaLong}
\be
   \p_\mu \gam_A = {{\Gam_\mu}^B}_A \gam_B~,~~~~~
   \p_\mu s_{\tl A} = {{\Gam_\mu}^{\tl B}}_{\tl A} s_{\tl B}
\lbl{5.18}
\ee
from which we find\ci{PavsicKaluzaLong} that
 $\p_\mu \psi^B$ and
$\p_\mu \psi^{\tl B}$ must be replaced with the covariant derivatives
\be
  \DD_\mu \phi^B = \p_\mu \phi^B + \Gam_\mu^{BC} \phi_C~~~
{\rm and}~~~
  \DD_\mu \phi^{\tl B} = \p_\mu \phi^{\tl B} 
  + {\Gam_\mu}^{\tl B \tl C} \phi_{\tl C}.
\lbl{5.19}
\ee
Then, in particular, the position dependent $\gam^\mu$ gives curved metric
according to $\gam_\mu(x) \cdot \gam_\nu (x) = g_{\mu \nu}$.
In addition, one also needs to include a kinetic term for $g_{\mu \nu}$ or
the connection ${\Gam_\mu}^{BC}$ (or for ${\Gam_\mu}^{{\tl B}{\tl C}}$).

Alternatively, one can find a solution $\phi^A$
(or, equivalently $\psi^{\tl A}$)
of the flat space equation (\ref{5.14c}), with $\gam_\mu \cdot \gam_\nu
=\eta_{\mu \nu}$, and calculate the
expectation value $\langle \gam_\mu \rangle$ according to Eq.\,(\ref{5.2}),
and then obtain the metric
\be
  g_{\mu \nu} = \langle \gam_\mu \rangle \cdot \langle \gam_\nu \rangle
\lbl{5.20a}
\ee
of a curved spacetime, induced by the fields $\phi^A$. No kinetic term for the
field $g_{\mu \nu}(x)$ or the corresponding connection  is necessary in
such a procedure. A curved spacetime metric comes directly from the fields
$\phi^A$ (or $\psi^{\tl A}$) which are solutions of the flat space equation
(\ref{5.14c}).

In both procedures the metric is given in terms of the fields $\phi^A$
(or $\psi^{\tl A}$). Equating the metrics (\ref{5.13}) and (\ref{5.20a}),
we have
\be
  \p_\mu \phi^A \p_\nu \phi^B \eta_{AB} =
  \phi^A C_{A B \mu}^a \phi^B \phi^C C_{CD \nu}^b \phi^D \gam_a \gam_b.
\lbl{5.21}
\ee
Here we have used Eqs.\,(\ref{5.2})--(\ref{5.4}) in which we replaced
$\psi^{\tl A}$ with $\phi^A$, as suggested by (\ref{5.1}). Eq.\,(\ref{5.21}) is
a condition that the fields $\phi^A$ must satisfy. Such a condition can
be satisfied if we start from the action
\be
  I = \int \dd^4 \left ( \hf \p_\mu \phi^A \p_\nu \phi^B \eta_{AB}
  \eta^{\mu \nu} - \mbox{$\frac{1}{4!}$} \lambda_{ABCD} \phi^A \phi^B \phi^C
  \phi^D \right )
\lbl{5.22}
\ee
with a quartic self-interaction term. The equations of motion are then
\be
  \p_\mu \p^\mu \phi_A  + \frac{1}{3!} \lambda_{ABCD} \phi^B \phi^C \phi^D
  = 0,
\lbl{5.23}
\ee
from which we obtain
$$\int \dd^4 x \, \left (\p_\mu \phi^A \p^\mu \phi_A -
  \frac{1}{3!} \lambda_{ABCD} \phi^B \phi^C \phi^D \right )
 \hs{5cm}$$
\be
 \hs{2cm} =  - \int \dd^4 x \, \left ( \phi^A \p_\mu \p^\mu \phi_A +
  \frac{1}{3!} \lambda_{ABCD} \phi^B \phi^C \phi^D \right ) = 0.
\lbl{5.24}
\ee
The latter equation also comes from (\ref{5.21}) after contracting
with $\eta^{\mu \nu}$ and integrating over $x$, provided that we identify
\be
  \frac{1}{3!} \lambda_{ABCD} = C_{AB \mu}^a C_{CD \nu}^b \eta^{\mu \nu}
  \eta_{ab},
\lbl{5.25}
\ee
where $\eta_{ab}= \gam_a \cdot \gam_b$.

In the action (\ref{5.22}) we have yet another possible generalization of the
non interacting action (\ref{5.11}) (the other generalization was the ``minimal
surface" action (\ref{5.12})). We have thus arrived at a fascinating result
that the spacetime metric $g_{\mu \nu}$ can be induced by Clifford algebra
valued field $\phi^A \gam_A$ that satisfies the quartic action principle
(\ref{5.22}).

\subsection{Fermion creation operators, branes as vacuums, branes with holes,
and induced gravity}

The procedure described in Sec.\,\ref{sec5}.1 can be considered as a special case of
quantized fields (\ref{4.23}) at a fixed spacetime point $x$. We will
now start from a generic object of the form (\ref{4.23}). It consists of the
terms such as
\be
  \psi^{\mu_1(x_1) \mu_2(x_2)... \mu_r (x_r)} h_{\mu_1 (x_1)} h_{\mu_2(x_2)} ...
  h_{\mu_r (x_r)} \Omega,
\lbl{5.27}
\ee
where we assume that $\Omega$ is the vacuum given by Eq.\,(\ref{4.19}).
The operator $h_{\mu_i (x_i)}$ creates a fermion at a point $x_i$. The
product of operators $h_{\mu_i (x_i)}h_{\mu_j (x_j)}$ creates a fermion
at $x_i$ and another fermion at $x_j$. By a generic expression (\ref{5.27})
we can form any structure of fermions, e.g., a spin network. In the limit
in which there are infinitely many densely packed fermions, we obtain arbitrary extended
objects, such as strings, membranes, $p$-branes, or even more general objects,
including instantonic branes, considered in Sec.\,\ref{sec2}.

Let us use the following compact notation for a state of many fermions
forming an extended object in spacetime:
\be
   \left ( \prod_{\mu,x \in {\cal R}} h_{\mu (x)} \right ) \Omega.
\lbl{5.28}
\ee
Here the product runs over spacetime points $x\in {\cal R}$ of a region
${\cal R}$ of spacetime $M_D$. In particular, ${\cal R}$ can be
a $p$-brane's world sheet $V_{p+1}$, whose parametric equation is
$x^\mu = X^\mu (\sigma^a)$, $\mu =0,1,2,...,D-1$, $a = 1,2,...,p+1$, or it
can be a brane-like instantonic object, also described by some functions
$X^\mu (\sigma)$.
Then the product of operators in Eq.\,(\ref{5.28}) can be written in the
form
\be
   \prod_{\mu,x=X(\sigma)} h_{\mu (x)} \equiv h [X^\mu (\sigma)],
\lbl{5.29}
\ee
where $h[X^\mu (\sigma)]$ is the operator that creates a brane or an
instantonic brane (that we will also call ``brane").
Here a brane is an extended objects consisting of infinitely many
fermions, created according to
\be
  \psi_{\rm brane} = h[X^\mu (\sigma)] \Omega 
  = \left (\prod_{\mu,x=X(\sigma)} h_{\mu (x)} \right ) \Omega
\lbl{5.30}
\ee

To make contact with the usual notation, we identify
\be
  \Omega \equiv |0 \rangle ~,~~~~~ 
  h[X^\mu (\sigma)] \equiv b^\dg [X^\mu (\sigma)]~,~~~~~\psi_{\rm brane}
  \equiv |X^\mu (\sigma) \rangle,
\lbl{5.31}
\ee
and write
\be
  |X^\mu (\sigma) \rangle = b^\dg [X^\mu (\sigma)] |0 \rangle .
\lbl{5.32}
\ee
A generic single brane state is a superposition of the brane states:
\be
  |\Psi \rangle = \int |X^\mu (\sigma) \rangle {\cal D} X (\sigma) 
  \langle X^\mu (\sigma)| \Psi \rangle .
\lbl{5.33}
\ee
In the notation of Eqs.\,(\ref{5.27})--(\ref{5.30}), the latter expression
reads
\be
  \Psi = \int {\cal D} X (\sigma) \psi [X^\mu (\sigma)] h [X^\mu (\sigma)]
  \Omega,
\lbl{5.33a}
\ee
where
\be
   \psi [X^\mu (\sigma)] = \lim_{r \rightarrow \infty,\Delta x_i \rightarrow 0}
   \psi^{\mu_1 (x_1)...\mu_r (x_r)} .
\lbl{5.33b}
\ee

However, besides single brane states, there are also two-brane, three-brane,
and in general, many-brane states. The brane Fock-space states are thus
\be
  b^\dg [X_1^\mu (\sigma)] |0 \rangle~,~~~
  b^\dg [X_1^\mu (\sigma)]  b^\dg [X_2^\mu (\sigma)]|0 \rangle~,~~~
  b^\dg [X_1^\mu (\sigma)]... b^\dg [X_r^\mu (\sigma)]|0 \rangle~,...
\lbl{5.34}
\ee
A generic brane state is a superposition of those states.

If we act on the brane state (\ref{5.30}) with the operator $\bh_{\mu' (x')}$,
we have
\be
  \bh_{\mu' (x')} \Psi_{\rm brane} = \bh_{\mu' (x')}
  \left (\prod_{\mu,x=X(\sigma)} h_{\mu (x)} \right ) \Omega .
\lbl{5.35}
\ee
If $x'$ is outside the brane, then nothing happens. But is $x'$ is a position
on the brane, then (\ref{5.35}) is a a state in which the particle at
$x'$ with the spin orientation $\mu$ is missing. In other words, (\ref{5.35})
is a brane state with a hole at $x'$.

We may also form two hole state, many-hole states, and the states with
a continuous set of holes,
\be
  \left ( \prod_{\mu, x\in {\cal R}_1} \bh_{\mu (x)} \right ) \Psi_{\rm brane}
  =\left ( \prod_{\mu, x\in {\cal R}_1} \bh_{\mu (x)} \right )
  \left (\prod_{\mu,x=X(\sigma)} h_{\mu (x)} \right ) \Omega ,
\lbl{3.35b}
\ee
where ${\cal R}_1 \subset {\cal R} = \{X^\mu (\sigma)\}$. For instance,
${\cal R}_1$ can be a string or a brane of a lower dimensionality than
the brane $X^\mu (\sigma^a)$.

If the space into which the brane is embedded has many dimensions, e.g.,
$D=10>p+1$, then the brane's worldsheet $V_{p+1}$ can represent our
spacetime\footnote{For more details on how an instantonic brane is
related to our evolving spacetime, see
Refs.\,\ci{PavsicBook,PavsicTimeProblem}.}
 which, if $p+1>4$, has extra dimensions. The induced metric
on $V_{p+1}$ can be curved, and so we have curved spacetime. We have thus
arrived at the brane world scenario. Holes in the brane are particles.
More precisely, the point like holes in the worldsheet $V_{p+1}$ are
instantonic point particles, whereas the string like holes are instantonic
strings, which can be either space like or time like
(see Refs.\,\ci{PavsicBook,PavsicTimeProblem}).

Let me now outline how the induced metric on a brane $V_{p+1}$ could be
formally derived in terms of the operators $h_{\mu(x)}$, $\bh_{\mu(x)}$.
The corresponding operators in orthogonal basis are (see (\ref{4.12}),
(\ref{4.13})),
\bear
  &&h_{1 \mu (x)} = \frac{1}{\sqrt{2}}(h_{\mu (x)} + \bh_{\mu(x)}),
  \lbl{5.36}\\
  &&h_{2 \mu (x)} = \frac{1}{i\sqrt{2}}(h_{\mu (x)} - \bh_{\mu(x)}),
  \lbl{5.37}
\ear
satisfy the Clifford algebra relations
\be
  h_{i \mu(x)} \cdot h_{j \nu(x')} = \delta_{ij} \eta_{\mu \nu} \delta (x-x').
\lbl{5.38}
\ee
In particular,
\be
  h_{1 \mu(x)} \cdot h_{1 \nu(x)} = \eta_{\mu \nu} \delta(0).
\lbl{5.39}
\ee
Comparing the latter result with
\be
   \gam_\mu \cdot \gam_\nu = \eta_{\mu \nu},
\lbl{5.40}
\ee
we find that\footnote{
Such notation could be set into a rigorous form if, e.g.,
in Eq.\,(\ref{5.38}) we replace $\delta(x-x')$ with $\frac{1}{a \sqrt{\pi}}
{\rm exp} [-\frac{(x-x')^2}{a^2}]$ and $\delta(0)$ with
$``\delta(0)" \equiv \frac{1}{a \sqrt{\pi}}$. Then Eq.\,(\ref{5.39}) is replaced by
$h_{1 \mu(x)} \cdot h_{1 \nu(x)} = \eta_{\mu \nu} ``\delta(0)"$. By inserting into the
latter equation the relation $h_{1 \mu(x)} = \gam_\mu \sqrt{``\delta(0)"}$,
we obtain $\gam_\mu \cdot \gam_\nu = \eta_{\mu \nu}$, which also holds in the limit
$a \to 0$, because $``\delta(0)"$ has disappeared
from the equation.}
\be
  h_{1 \mu(x)} = \gam_{\mu} \sqrt{\delta(0)}.
\lbl{5.41}
\ee
This means that up to an infinite constant, $h_{1 \mu(x)}$ is
proportional to $\gam_\mu$, a basis vector of Minkowski spacetime. Thus,
a proper renormalization of $h_{1 \mu(x)}$ gives $\gam_\mu$.

In a given quantum state $\Psi$ we can calculate the expectation value
of $h_{i \mu(x)}$ according to
\be
  \langle h_{i \mu(x)} \rangle = \langle \Psi^\ddg h_{i \mu(x)} \Psi \rangle_1,
\lbl{5.42}
\ee
where the subscript 1 means vector part of the expression in the bracket.
The inner product gives the expectation value of the metric:
\be
  \langle \rho_{i \mu(x)j \nu(x')} \rangle 
  = \langle h_{i \mu(x)} \rangle \cdot \langle h_{j \nu(x')} \rangle .
\lbl{5.43}
\ee
This is the metric of an infinite dimensional manifold that, in general,
is curved. In Refs.\,\ci{PavsicBook}, a special case of such a manifold,
for $i=j=1$, called {\it membrane space} ${\cal M}$, was considered.
It was shown how to define connection and curvature of ${\cal M}$.

Taking $i=j=1$ and $x=x'$ in Eq.\,(\ref{5.43}), we have
\be \langle \rho_{1 \mu(x)\, 1 \nu(x)} \rangle 
  = \langle h_{1 \mu(x)} \rangle \cdot \langle h_{1 \nu(x)} \rangle .
\lbl{5.44}
\ee
Upon renormalization according to (\ref{5.41}) (see Footnote\, 8),
we obtain
\be
  \langle g_{\mu \nu} (x) \rangle 
  = \langle \gam_\mu (x) \cdot \langle \gam_{\nu}(x) \rangle,
\lbl{5.45}
\ee
where
\be
  \langle g_{\mu \nu} (x) \rangle 
  =\langle \rho_{1 \mu(x) 1\nu(x)} \rangle \frac{1}{\sqrt{``\delta (0)"}}
\lbl{5.46}
\ee
is a position dependent metric of spacetime. We expect that the corresponding
Riemann tensor is in general different from zero.

As an example let us consider the expectation value of a basis vector
$h_{1 \mu(x)}$ in the brane state (\ref{5.30}):
\be
  \langle h_{1 \mu(x)} \rangle = \langle \Psi_{\rm brane}^\ddg h_{1 \mu(x)}
  \Psi_{\rm brane} \rangle_1 =
 \langle \Psi_{\rm brane}^\ddg \frac{1}{\sqrt{2}} (h_{\mu(x)} +\bh_{\mu(x)})
 \Psi_{\rm brane} \rangle_1 .
\lbl{5.47}
\ee
From Eq.\,(\ref{5.35}) in which the vacuum $\Omega$ is defined according to
(\ref{4.19}), we have
\be
  \bh_{\mu (x)} \Psi_{\rm brane} = \begin{cases}
  \Psi_{\rm brane}({\check x}), & x\in\text{brane};\\[2mm]
   0, & x\not\in\text{brane}.
   \end{cases}
\lbl{5.48}
\ee
Here $\Psi_{\rm brane}({\check x})$, with the accent ``\ {\large \v \ }\ ''
on $x$, denotes the brane with a hole at $x$.
The notation $x\in\text{brane}$ means that $x$ is on the brane, whereas
$x\not\in\text{brane}$ means that $x$ is outside the brane created according to
(\ref{5.30}).

Because $( \bh_{\mu (x)} \Psi_{\rm brane})^\ddg 
= \Psi_{\rm brane}^\ddg h_{\mu (x)}$, we also have
\be
  \Psi_{\rm brane}^\ddg h_{\mu (x)} = \begin{cases}
  \Psi_{\rm brane}({\check x})^\ddg, & x\in \text{brane};\\[2mm]
   0, & x\not\in\text{brane}. \\
   \end{cases}
\lbl{5.49}
\ee
For the expectation value of $h_{1 \mu (x)}$ we then obtain
\be
  \langle h_{1 \mu (x)} \rangle = \begin{cases}
  \frac{1}{\sqrt{2}} \langle \Psi_{\rm brane}({\check x})^\ddg \Psi_{\rm brane}
  \rangle_1 + 
  \frac{1}{\sqrt{2}} \langle \Psi_{\rm brane}^\ddg \Psi_{\rm brane}(\text{\v x})
  \rangle_1, & x\in\text{brane};\\[2mm]
   0, & x\not\in\text{brane}. \\
   \end{cases}
\lbl{5.50}
\ee
A similar expression we obtain for $\langle h_{2 \mu(x)}$.
The expectation value of the metric\footnote{
Notice that the expectation value of the metric is {\it not} defined as
$\langle \rho_{i \mu(x) j \nu(x')} \rangle 
= \langle \Psi^\ddg \rho_{i \mu(x) j \nu(x')} \Psi \rangle$, but as
$\langle \rho_{i \mu(x) j \nu(x')} \rangle=
\langle h_{i \mu (x)} \rangle \cdot \langle h_{j \nu (x)} \rangle$.}
(\ref{5.43}) is
\be
  \langle \rho_{i \mu (x) j \nu (x')} \rangle = \begin{cases}
  \langle \rho_{i \mu(x) j \nu(x')} \rangle|_{\rm brane}, & 
  \text{on the brane};\\[2mm]
   0, & \text{outside the brane}. \\
   \end{cases}
\lbl{5.51}
\ee
An interesting result is that outside the brane the expectation value
of the metric is zero. Outside the brane, there is just the vacuum $\Omega$.
The expectation value of a vector $h_{i \mu (x)}$ in the vacuum, given by
(\ref{4.19}), is zero, and so is the expectation value
$\langle \rho_{i \mu(x) j \nu(x')} \rangle$. This makes sense, because
the vacuum $\Omega$ has no orientation that could be associated with a
non vanishing effective vector. In $\Omega$ there also are no special points
that could determine distances, and thus a metric.
This is in agreement with the concept of configuration space, developed in
Refs.\,\ci{PavsicBook}, (see also Sec.\,\ref{sec2}), according to which outside
a configuration there is no space and thus no metric: a physical space is
associated with configurations, e.g., a system of particles, branes, etc.;
without a configuration there is no  physical space. In other words,
a concept of a physical space unrelated to a configuration of physical
objects has no meaning. Our intuitive believing that there exists
a three (four) dimensional space(time) in which objects live is deceiving us.
The three (four) dimensional space(time) is merely a subspace of the
multidimensional configuration space of our universe, in which only position
of a single particle is allowed to vary, while positions of all remaining
objects are considered as fixed. Of course this is only an idealization.
In reality, other objects are not fixed, and we have to take into account,
when describing the universe, their configuration subspaces as well.
Special and general relativity in 4-dimensional spacetime is thus a
special case of a more general relativity in configuration space.
Quantization of general relativity has failed, because it has not taken
into account the concept of configuration space, and has not recognized
that 4D spacetime is a subspace of the huge configuration space associated
with our universe. The approach with quantized fields presented in this
work has straightforwardly led us to the concept of many particle
configurations and effective curved spaces associated with them.

If in Eq.\,(\ref{5.51}) we take $i=j$, $x=x'$, and use
Eqs.\,(\ref{5.44}--(\ref{5.46}), then we obtain
\be
  \langle g_{\mu \nu} (x) \rangle = \begin{cases}
  g_{\mu \nu} (x) |_{\rm brane} \neq 0, & 
  \text{on the brane};\\[2mm]
   0, & \text{outside the brane}. \\
   \end{cases}
\lbl{5.52}
\ee
It is reasonable to expect that detailed calculations will give the result
that $g_{\mu \nu} (x) |_{\rm brane}$ is the induced metric on the brane, i.e.,
\be
  g_{\mu \nu} (x) |_{\rm brane} = \p_a X^\mu \p_b X^\nu \eta_{\mu \nu}
  \equiv f_{ab}.
\lbl{5.53}
\ee
Recall that the brane can be our spacetime. We have thus pointed to a possible
derivation of a curved spacetime metric from quantized fields
in higher dimensions.

\section{Quantized fields and Clifford space}\lbl{sec6}

In the previous section we considered fermion states that are generated by
the action of creation operators on the vacuum $\Omega$ according to
Eq.\,(\ref{5.27}). In particular, a many fermion state can be a brane,
formed  according to eq.\,(\ref{5.28}). In Sec.\,\ref{sec2} we showed that a brane
can be approximately described by a polyvector (\ref{2.5})
(see also (\ref{2.6})), which is a superposition of the Clifford algebra
basis elements
\be
  {\ul 1},~~ \gam_\mu \wg \gam_\nu,~~\gam_\mu \wg \gam_\nu \wg \gam_\rho,~~
  \gam_\mu \wg \gam_\nu \wg \gam_\rho \wg \gam_\sigma .
\lbl{6.1}
\ee
This means that a Fock space element of the form (\ref{5.28}) can be
mapped into a polyvector:
\be
  \left ( \prod_{\mu,x \in {\cal R}} h_{\mu(x)} \right ) \Omega \longrightarrow
  x^M \gam_M .
\lbl{6.2}
\ee

As an example let us consider the case in which the region ${\cal R}$ of
spacetime is a closed line, i.e., a loop. The holographic projections
of the area enclosed by the loop are given in terms of the bivector
coordinates $X^{\mu \nu}$. The loop itsel is described\footnote{
Of course, there is a class of loops, all having
the same $X^{\mu \nu}$.}
by a bivectors $X^{\mu \nu} \gam_\mu \wg \gam_\nu$. So we have the mapping
\be
  \left ( \prod_{\mu,x \in {\rm loop}} h_{\mu(x)} \right ) \Omega
   \longrightarrow x^{\mu \nu} \gam_{\mu \nu} .
\lbl{6.3}
\ee

With the definite quantum states, described by Eq.\,(\ref{5.28}) or
(\ref{5.30}) (see also (\ref{5.32})), which are the brane basis states,
analogous to position states in the usual quantum mechanics, we can form a
superposition (\ref{5.33a}) (see also (\ref{5.33})). To such an indefinite
brane state there corresponds a state with indefinite polyvector
coordinate $X^M$:
\be
  \int {\cal D} X(\sigma) \Psi [X(\sigma)] h[X(\sigma)] \Omega
  \longrightarrow \phi (x^M) .
\lbl{6.4}
\ee

In particular, if $h[X(\sigma)] \Omega$ is a loop, then we have the mapping
\be
 \int {\cal D} X(\sigma) \Psi [X(\sigma)] h[X(\sigma)] \Omega
  \longrightarrow \phi (x^{\mu \nu}) .
\lbl{6.5}
\ee

The circle is thus closed. With the mapping (\ref{6.2}) we have again arrived
at the polyvector $x^M \gam_M$ introduced in Sec.\,\ref{sec2}.
The polyvector coordinates $x^M$ of a classical system satisfy the dynamics
as formulated in Refs.\,\ci{PavsicKaluzaLong,PavsicArena,PavsicMaxwellBrane}.
That dynamics can be generalized to
super phase space as discussed in Sec.\,\ref{sec3}, where besides the commuting
coordinates $x^\mu$, $\mu=0,1,2,3$, we introduced the Grassmann coordinates
$\xi^\mu$.
In the quantized theory, the wave function $\psi(x^\mu,\xi^\mu)$ represents
a 16-component field, $\phi^A$, $A=1,2,...,16$, that depends on position
$x^\mu$ in spacetime, and
satisfies the Dirac equation (\ref{5.14b}) and the multicomponent Klein-Gordon
equation (\ref{5.14a}). In analogous way, besides commuting polyvector
coordinates $x^M$, $M=1,2,...,16$, we 
have the corresponding Grassmann coordinates $\xi^M$, and the wave function
$\phi(x^M)$ is generalized to $\phi(x^M,\xi^M)$. The expansion of
$\phi (x^M,\xi^M)$ in terms of $\xi^M$ gives a $2^{16}$-component field,
$\phi^A$, $A=1,2,...,2^{16}$,
that depends on position $x^M$ in Clifford space, and satisfies the
generalized Dirac equation, $\gam^M \p_M \phi^A (x^M)$.

As the evolution
parameter, i.e., the time along which the wave function evolves, we can
take the time like coordinate $x^0$, or the time-like coordinate $\sigma$.
Alternatively, we can take the light-like coordinate $s$, defined
in Eq.\,(\ref{2.14}), as the evolution parameter. Then, as shown in Ref.\,
(\ci{PavsicLocalTachyons}), the Cauchy problem can be well posed, in
spite of the fact that in Clifford space there are eight time-like dimensions,
besides eight space-like dimensions.
Moreover, according to Refs.\,\ci{Woodard,PavsicPseudoHarm,PavsicSaasFee},
there are no ghosts in such spaces, if the theory is properly quantized, and
in Refs.\,\ci{PUStable}--\ci{PUdamp} it was shown that the stability of solutions
can be achieved even in the presence of interactions.

We can now develop a theory of such quantized fields in Clifford space
along similar lines as we did in Secs.\,\ref{sec4} and \ref{sec5} for
the quantized fields in the ordinary spacetime. So we can consider the analog
of Eqs.\,(\ref{5.28})--(\ref{5.53}) and arrive at the induced metric on a
4-dimensional surface $V_4$ embedded in the 16-dimensional Clifford space.
Whereas in Eqs.\,(\ref{5.28})--(\ref{5.53}) we had hoc postulated the
existence of extra dimensions, we now see that extra dimensions are
incorporated in the configuration space of brane like objects created by
the fermionic field operators $h_{\mu (x)}$. Our spacetime can thus be a
curved surface embedded in such a configuration space.

\section{Conclusion}\lbl{sec7}

Clifford algebras are very useful to describe extended objects as points in
Clifford spaces, which are subspaces of configuration spaces. The Stueckelberg
evolution parameter can be associated with the scalar and the pseudoscalar
coordinate of the Clifford space.

The generators of orthogonal and symplectic Clifford algebras. i.e.,
the orthogonal and symplectic basis vectors, behave,
respectively, as fermions and bosons. Quantization of a classical theory
is the shift of description from components to the (orthogonal or symplectic)
basis vectors.

We have found that a natural space to start from is a phase space, which can be
either orthogonal or symplectic. We united both those phase spaces into
a super phase space, whose points are described by
anticommuting (Grassmann) and commuting coordinates, the basis vectors
being the generators of orthogonal and symplecting Clifford algebras.
We have considered the Clifford algebra $Cl(8)$ constructed
over the 8-dimensional orthogonal part of the super phase space. Remarkably,
the 256 spinor states of $Cl(8)$ can be associated with all the particles of
the Standard Model,  as well as with additional particles that do not
interact with our photons and are therefore invisible to us. This model thus
predicts dark matter. Moreover, it appears to be a promising step towards
the unification of elementary particles and interactions
(see also\,\ci{PavsicKaluza,PavsicKaluzaLong,PavsicMaxwellBrane}.

Both, orthogonal and symplectic Clifford algebras can be generalized
to infinite dimensions, in which case their generators (basis vectors)
are bosonic
and fermionic field creation and annihilation operators. In the Clifford
algebra approach to field theories, a vacuum is the product of infinite,
uncountable number of Fermionic field creation operators. They can form many
sorts of possible vacuums as the seas composed of those field operators.
In particular, strings and branes can be envisaged as being such seas. The
field operators, acting on such brane states, can create holes in the branes,
that behave as particles. From the expectation values of vector operators
in such a one, two, or many holes brane state, we can calculate the metric
on the brane. According to the brane world scenario, a brane can be our
world. We have found that holes in a fermionic brane behave as particles,
i.e., matter, in our world, and that the metric on the brane can be
quantum mechanically induced by means of the fermionic creation and annihilation
operators. We have thus found a road to quantum gravity that seems to avoid the
usual obstacles.

\hs{4mm}

\centerline{\bf Acknowledgement}

This work has been supported by the Slovenian Research Agency.

\end{document}